\documentclass[11pt,a4paper]{article}
\usepackage{float}
\usepackage[margin=1in]{geometry}
\usepackage{times}
\usepackage[authoryear,round]{natbib}

\usepackage{amsmath,amssymb,amsfonts}
\usepackage{graphicx}
\usepackage{textcomp}
\usepackage{xcolor}
\usepackage{booktabs}
\usepackage{subcaption}

\usepackage{svg}
\setsvg{inkscapelatex=true}

\usepackage[pdfencoding=auto,psdextra]{hyperref}
\usepackage{url}
\usepackage{multirow}
\usepackage{makecell}
\usepackage{tikz}
\usetikzlibrary{matrix}
\usepackage{glossaries}
\usepackage{amsthm}

\usepackage{abstract}

\newcommand{\node}{\mathcal{N}}

\DeclareMathOperator*{\argmin}{arg\,min}

\begin{document}

\title{HypeR Adaptivity: Joint $hr$-Adaptive Meshing via Hypergraph Multi-Agent Deep Reinforcement Learning}

\author{
Niccolò Grillo$^{1}$,
James Rowbottom$^{2}$,
Pietro Liò$^{1}$,
Carola-Bibiane Schönlieb$^{2}$,
Stefania Fresca$^{3}$
\\[1ex]
$^{1}$Department of Computer Science and Technology, University of Cambridge, UK \\
$^{2}$Department of Applied Mathematics and Theoretical Physics, University of Cambridge, UK \\
$^{3}$Department of Mechanical Engineering, University of Washington, Seattle, WA, USA
}

\date{}

\maketitle

\begin{abstract}
 Adaptive mesh refinement is central to the efficient solution of partial differential equations (PDEs) by means of the finite element method (FEM). Classical $r$-adaptivity optimizes vertex positions but requires solving expensive auxiliary PDEs such as the Monge-Ampère equation, while classical $h$-adaptivity modifies topology through element subdivision but suffers from computationally expensive error indicator computation and remains fundamentally constrained by isotropic refinement patterns that impose accuracy ceilings. Combined $hr$-adaptive techniques naturally outperform single-modality approaches, yet inherit both computational bottlenecks and the restricted cost-accuracy trade-off. An emerging field of machine learning methods for adaptive mesh refinement has sought to overcome these limitations, but existing approaches address $h$-adaptivity or $r$-adaptivity in isolation. In this respect, we present HypeR, a deep reinforcement learning framework that jointly optimizes mesh relocation and refinement in a unified formulation. HypeR casts the joint adaptation problem using tools from hypergraph neural networks and multi-agent reinforcement learning. Refinement is formulated as a heterogeneous multi-agent Markov decision process (MDP) where a swarm of element agents decide discrete refinement actions, while relocation follows an anisotropic diffusion-based policy operating on vertex agents with provable prevention of mesh tangling. The reward function combines local and global error reduction to promote general accuracy. Across benchmark PDEs, HypeR reduces approximation error by up to 6--10$\times$ versus state-of-art $h$-adaptive baselines at comparable element counts, breaking through the uniform refinement accuracy ceiling that constrains subdivision-only methods. The framework produces meshes with improved shape metrics and alignment to solution anisotropy, demonstrating that jointly learned $hr$-adaptivity strategies can substantially enhance the capabilities of automated mesh generation.
\end{abstract}


\maketitle


\section{Introduction}
Partial differential equations (PDEs) model a vast array of physical phenomena, from fluid dynamics and heat transfer to structural mechanics and electromagnetic fields~\cite{evans2010partial,leveque2007finite}. The finite element method (FEM) has emerged as the predominant numerical method for solving PDEs at scale, offering robustness, well-established error bounds, and mature software implementations~\cite{brenner2008mathematical,ern2004theory}. However, even with highly optimized solvers, the computational cost of simulations grows rapidly with the number of {degrees of freedom (DOFs)}. Denoting by $N$ the total number of DOFs, memory requirements typically grow superlinearly, and the cost of direct sparse factorization scales as $\mathcal{O}(N^{3/2})$ in two dimensions and as $\mathcal{O}(N^{2})$ in three dimensions. By contrast, optimal multigrid or multilevel iterative methods can in many cases achieve $\mathcal{O}(N)$ complexity~\cite{saad2003iterative}. This makes large-scale problems computationally prohibitive~\cite{babuska1994p}, motivating mesh adaptation: strategically redistributing mesh resolution (i.e., the local density of DOFs or mesh vertices) that optimizes the cost-accuracy trade-off in order to capture solution features where they matter most.

Classical mesh adaptation strategies include two complementary techniques, each addressing different aspects of the accuracy-efficiency trade-off. $h$-adaptation modifies mesh topology through element subdivision or coarsening, dynamically adjusting local resolution based on a posteriori error estimators. Popular approaches include the Zienkiewicz-Zhu (ZZ) superconvergent patch recovery technique~\cite{zienkiewicz1992superconvergent}, which estimates element-wise errors $\eta_K$ through gradient reconstruction, and Dörfler marking strategies~\cite{dorfler1996convergent}, which refine elements contributing to a fixed fraction of the total error~\cite{verfurth1996review,ainsworth2000posteriori}. In contrast, $r$-adaptation relocates a fixed number of mesh vertices while preserving connectivity, optimizing vertices positions through a continuous mapping $\Phi$. 
This typically requires solving auxiliary PDE problems, such as the Monge-Ampère equation for optimal transport-based mesh movement~\cite{budd2009adaptivity,delzanno2008optimal,huang2011adaptive}. While $h$-methods can arbitrarily increase resolution, they suffer from exponential growth in DOFs. In contrast, $r$-methods maintain a fixed number of DOFs, but are limited by the initial mesh topology. The optimal strategy - $hr$-adaptation - combines both approaches, yet traditional numerical methods treat them sequentially, missing potential synergies and requiring complex bookkeeping of element hierarchies and projection operators~\cite{alauzet2016decade,bank1983some}. Crucially, these combined techniques inherit the computational bottlenecks of both modalities, from solving expensive auxiliary PDEs to costly error indicator computations, limiting their practical efficiency. \\

An emerging field of machine learning methods has sought to overcome these computational bottlenecks, but existing approaches have largely addressed $h$-adaptivity or $r$-adaptivity in isolation. For $h$-adaptation, deep reinforcement learning (RL) offers a fundamentally different paradigm. By formulating mesh adaptation as a Markov Decision Process (MDP) with state space $\mathcal{S}$ (mesh configurations), action space $\mathcal{A}$ (adaptation operations), and reward function $R$ (error reduction), RL agents can learn policies $\pi: \mathcal{S} \to \mathcal{A}$ that optimize long-term error-efficiency trade-offs~\cite{andrew2018reinforcement}. This approach has shown promise, with methods like ASMR++~\cite{freymuth2024asmrplus} employing swarm-based multi-agent RL to achieve order of magnitude speedups over uniform refinement. In parallel, graph neural network (GNN) approaches have targeted $r$-adaptation. Many of these, such as the M2N~\cite{song2022m2n} and UM2N~\cite{zhang2024towards} families, are trained in a supervised manner to imitate the complex mappings generated by classical auxiliary PDE solvers. A distinct GNN-based strategy, employed by G-Adaptivity~\cite{rowbottom2025gadaptivity}, bypasses these expensive targets and instead learns to directly minimize FEM error through differentiable simulation. However, each of these methods addresses only one aspect of mesh adaptation, inheriting fundamental limitations: $h$-only methods cannot exceed uniform refinement accuracy on the initial topology, while $r$-only methods cannot add DOFs for emerging features. \\

\begin{figure}[t]
  \centering
  \includegraphics[width=1\linewidth]{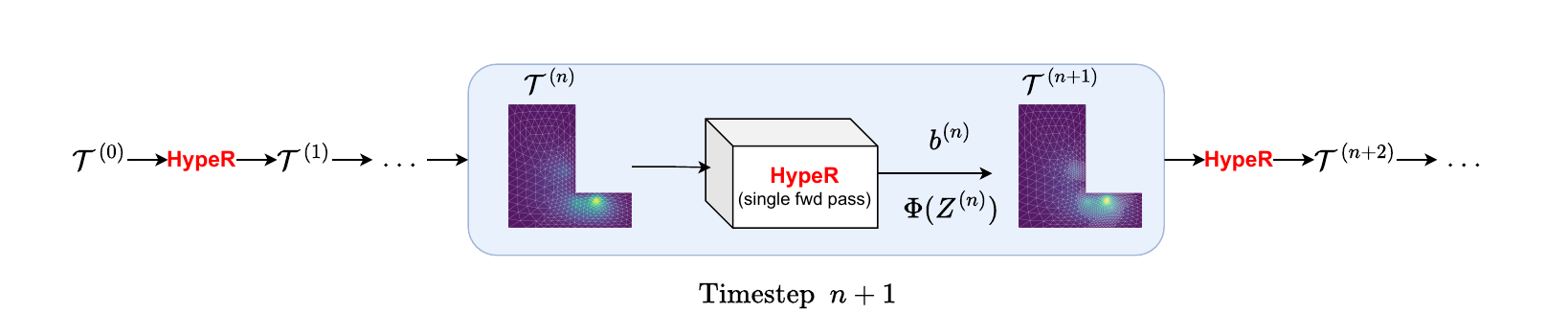}
\caption{A high-level schematic of the HypeR framework at inference. Starting from an initial coarse mesh $\mathcal{T}^{(0)}$, the HypeR policy network is applied iteratively. At each step, the network takes the current mesh $\mathcal{T}^{(n)}$ and, in a single forward pass, jointly outputs the refinement vector $b^{(n)}$ and the vertex relocation map $\Phi(\mathcal{Z}^{(n)})$ to produce the next adapted mesh $\mathcal{T}^{(n+1)}$.}
\label{fig:hypeR_inference_loop}
\end{figure}

In this work, we present HypeR, a deep RL framework that computes relocation and refinement actions in a single forward pass of a hypergraph neural network based rollout strategy. This process, visualized in Figure~\ref{fig:hypeR_inference_loop}, is applied iteratively starting from a coarse mesh. In each adaptation step, the policy network jointly determines the relocation map $\Phi$ through continuous vertex actions and the refinement vector $b$ through discrete element actions. This unified formulation directly optimizes the coupled pair $(\Phi,b)$ without resorting to staged procedures where intermediate calculations of the solution are needed, allowing relocation and refinement to inform one another within the same decision process.

HypeR is based on three interconnected principles. First, we represent the mesh as a hypergraph $\mathcal{H}^{(n)} = (\mathcal{Z}^{(n)}, \{K_j^{(n)}\}_j, H^{(n)}, A^{(n)})$, where vertices and element embeddings, i.e., $\mathcal{Z}^{(n)}$ and $\{K_j^{(n)}\}_j$, are linked via incidence and adjacency operators. In particular, hypergraph message passing~\cite{yadati2019hypergraph} 
{enables bidirectional exchange between the hyper-edges, which represent mesh elements, where agents make decisions of mesh refinement that changes local mesh topology, and vertices where agents make geometric decisions via the movement of each node. As excessive movement incurs a small cost but can strongly impact mesh quality and excessive refinement negates the need for mesh movement at a high computational cost, the coupling ensures that refinement decisions depend on local mesh quality and that relocation anticipates forthcoming subdivisions.}
Second, we formulate adaptation as a heterogeneous multi-agent Markov Decision Process (MDP) following the swarm RL paradigm~\cite{sosic2017inverse,huttenrauch2019deep}: vertex agents parameterize continuous displacements sampled from Gaussian policies, while element agents sample discrete refinement flags from Bernoulli policies. Third, we guarantee non-tangling relocation: the vertex policy is parameterized through a diffusion-based operator that preserves element orientation by ensuring $\det(J_\Phi|_K) > 0$, for all elements $K$. This construction prevents inversion while remaining differentiable, providing the orientation-preserving property required for valid FEM computations.

The synergy between the different components enables HypeR to overcome the limitations of existing approaches. More precisely, the hypergraph representation allows element agents to communicate refinement demands to vertex agents, which can relocate in advance vertices so that new elements are well-conditioned before subdivision. Conversely, regions of geometric distortion detected by vertex agents can trigger topological updates, guiding refinement where relocation alone is insufficient. This bidirectional coordination is learned through multi-objective Proximal Policy Optimization (PPO) ~\cite{schulman2017proximal} with separate policy and value heads for each agent type, producing meshes unattainable by pure $h$-adaptive or $r$-adaptive methods. Beyond error reduction, HypeR demonstrably improves mesh quality: elements exhibit less acute minimum angles, reduced skewness, and smoother gradation. In this way, the resulting anisotropy better aligns element orientation with solution features. These properties, quantified through standard shape metrics, directly enhance stability and approximation accuracy by reducing stiffness matrix conditioning issues and capturing directional features of the PDE solutions. Across benchmark PDEs including Poisson, heat diffusion, Stokes flow, and linear elasticity, HypeR reduces approximation error by up to $6-10\times$ compared to state-of-art $h$-adaptive methods at comparable element counts. Furthermore, the learned policies demonstrate robust zero-shot generalization, successfully adapting to unseen domain geometries and scales significantly larger than those seen during training. This establishes joint reinforcement-learned $hr$-adaptivity as a unified paradigm for automated mesh generation, achieving lower error and demonstrably higher mesh quality, as measured by classical shape metrics, and alignment with solution fields.

In this study, we validate the HypeR framework primarily on two-dimensional stationary problems using meshes with linear basis functions. However, the underlying formulation—specifically the hypergraph representation and multi-agent reinforcement learning architecture—is inherently dimension-agnostic. As we discuss in the methodology, extending HypeR to three-dimensional domains or higher-order elements requires little to no structural adjustment to the learning algorithm.

\subsection{Related Works}
In this section we present and review related research ranging from traditional mesh adaptation techniques to most recent machine learning-based methods. Classical methods provide rigorous foundations for $h$-adaptivity through a posteriori error estimators and marking strategies, and for $r$-adaptivity through PDE-based and variational relocation schemes. Most recently, learning-based methods framing refinement as a RL problem, relocation as a neural mesh movement task, or mesh generation as a supervised prediction problem, have emerged. These complementary directions, detailed below, define the current landscape, yet none provide a unified learning framework for $hr$-adaptivity, which is the focus of HypeR.

\paragraph{Traditional mesh adaptation methods.} Classical mesh adaptation separates $h$-methods (element refinement inducing topology modification) from $r$-methods (vertex relocation entailing geometry modification). Zienkiewicz-Zhu error estimators~\cite{zienkiewicz1992superconvergent} and Dörfler marking~\cite{dorfler1996convergent} represent $h$-adaptivity foundations with proven convergence rates~\cite{binev2004adaptive,stevenson2008optimality}. In contrast, $r$-adaptivity relies on different numerical schemes to reposition a fixed set of vertices, such as solving auxiliary PDEs (e.g., the Monge-Ampère equation~\cite{budd2009adaptivity,delzanno2008optimal,huang2011adaptive}), optimizing mesh quality via metric-based methods~\cite{yano2012optimization}, or minimizing a cost functional through variational approaches~\cite{mcrae2018optimal}. Combined $hr$-adaptivity strategies exist primarily in specialized applications~\cite{nagarajan2018review} using heuristic coordination. These methods rely on hand-crafted rules, limiting their ability to optimize the accuracy-cost trade-offs in complex scenarios.

\paragraph{Reinforcement learning for $h$-adaptivity.} Recent advances in RL have addressed adaptive mesh refinement through various sequential decision-making approaches. ~\cite{freymuth2023swarm} introduced swarm-based multi-agent RL (ASMR) where mesh elements act as collaborating agents that can split and therefore increase local mesh density, achieving 2 orders of magnitude speedup over uniform refinement. Its extension ASMR++~\cite{freymuth2024asmrplus} improves upon the previous approach with local rewards based on maximum error reduction, normalized agent mapping for stable reward propagation, and adaptive element penalties enabling multi-resolution policies without retraining. These features enable to overcome fundamental limitations in earlier RL approaches: VDGN~\cite{yang2023multiagent} used value decomposition networks for credit assignment but struggled with training stability on larger meshes, single-agent methods~\cite{yang2023reinforcement} suffered from prohibitive inference runtime by refining one element per timestep, and sweep-based approaches~\cite{foucart2023deep} created training-inference misalignment through random sequential sampling during training versus parallel execution during inference. DDynAMO~\cite{dzanic2024dynamo} extends key ideas from multi-agent RL-based $h$-adaptive mesh refinement to time-dependent problems through anticipatory refinement, but remains constrained by its reliance on structured 2D meshes, fixed observation windows, and PDE-specific features, limiting its scalability and general applicability. Other specialized approaches target specific domains like GMR-Net~\cite{kim2023gmrnet} for elliptic PDEs, Flow2Mesh~\cite{yu2024flow2mesh} for fluid dynamics, and recurrent networks~\cite{bohn2021recurrent} for marking strategies. Despite impressive results, all these $h$-adaptive methods are fundamentally limited to uniform refinement on the initial mesh structure.

{\paragraph{Neural approaches to $r$-adaptivity.} Recent approaches for mesh movement have investigated the use of GNNs to overcome the bottleneck of classical velocity-based methods relying on the costly solution of the Monge-Ampère equation. They all employ graph learning-based algorithms to relocate mesh vertices while maintaining fixed connectivity. In this regard, a series of mesh movement net
ùworks from the M2N family represent progressive extensions. In particular,  M2N~\cite{song2022m2n} introduced end-to-end mesh movement with neural splines and Graph Attention Networks trained on pre-generated Monge-Ampère meshes. UM2N~\cite{zhang2024towards} enabled zero-shot generalization across PDE types, relying only on the provision of a suitable monitor function and using graph transformer encoders, and UGM2N~\cite{wang2025ugm2n} eliminated the dependency on pre-adapted meshes through unsupervised learning with localized geometric features. ~\cite{rowbottom2025gadaptivity} proposed G-Adaptivity: a GNN diffusion-based vertex relocation framework which directly minimizes FEM error, through differentiable FEM solving, and preserves mesh validity via the diffusion-based formulation. Moreover, recent works include data-free approaches~\cite{hu2024better} and alternative diffusion-based frameworks~\cite{yu2025para2mesh}. All these approaches achieve 10-100× speedups over classical methods, however, these $r$-adaptive methods operate only on fixed connectivity, preventing resolution enhancement through refinement, which represents a fundamental limitation for evolving features or singularities.}

{\paragraph{Supervised learning for mesh generation.} Adaptive meshing by expert reconstruction (AMBER)~\cite{freymuth2025amber} introduced a
multi-scale hierarchical GNN trained with an experience replay buffer to predict the sizing field extracted from human expert generated meshes. The sizing field is then injected into standard full mesh generation algorithms for use in downstream tasks. GraphMesh~\cite{khan2024graphmesh} utilsies a GCN based architecture to predict  error distributions on meshes for general polygonal domains utilising domain aware nodes features in the form of mean value coordinates, while earlier works, like MeshingNet~\cite{zhang2020meshingnet} and its 3D extension~\cite{zhang2021meshingnet3d}, learn from a posteriori error estimate. Finally, image-based methods~\cite{huang2021machine} using CNNs on discretized domains struggle with varying resolution requirements. These supervised approaches require expert data or specific error estimates, limiting their generalization and applicability compared to RL methods.}

\subsection{Main Contributions}

This work introduces HypeR, the first RL framework for unified $hr$-adaptivity. Our main contributions are:

\begin{itemize}
    \item \textbf{Joint ${hr}$-adaptive RL formulation.} We present a framework that learns simultaneous mesh relocation and refinement policies through a dual-swarm MDP. Unlike prior works, restricted to either $h$-adaptivity (ASMR++, VDGN) or $r$-adaptivity (M2N, UM2N, G-Adaptivity), our approach discovers coordinated strategies that exploit the synergy between topology and geometry modifications.
    
    \item \textbf{Hypergraph neural architecture with heterogeneous message passing.} We introduce a novel mesh representation that treats vertices and elements as distinct node types in a hypergraph, enabling specialized information flow between geometric and topological levels. This architecture captures this duality in mesh adaptation, through separate but coupled embedding spaces connected by hypergraph convolutions.
    
    \item \textbf{Provably valid mesh operations through diffusion-based relocation.} We integrate the diffusion-based movement guarantees from G-Adaptivity into a learned policy framework, ensuring mesh validity. This bridges the gap between classical numerical guarantees and modern deep learning, providing the first RL approach to $r$-adaptivity that provably prevents mesh tangling, including non-convex domains.
    
  \item \textbf{Multi-objective optimization with distinct agent rewards.} We formulate mesh adaptation as a heterogeneous multi-agent problem where vertex and element agents optimize different objectives on different timescales. Vertex agents use short-horizon rewards ($\gamma_r \ll 1$) for immediate error reduction through relocation, while element agents employ long-horizon returns ($\gamma_h \approx 1$) to capture the forward impact of refinement decisions. Thanks to the differentiability of the RL loss, simultaneous gradient-based optimization of both refinement and relocation is possible.

    \item \textbf{Curriculum learning for coupled policy optimization.} We develop a two-phase training curriculum that addresses the non-stationary learning dynamics of joint adaptation. By initially training relocation with random refinement before enabling coordinated learning, we stabilize convergence and obtain robust policies that generalize to different PDE parameters and domain shapes and sizes.
    
    \item \textbf{Enhancement in adaptive mesh refinement accuracy.} HypeR achieves 6--10$\times$ error reduction compared to state-of-art $h$-adaptive methods at equivalent element counts across four benchmark PDEs. By overcoming the uniform refinement approximation limit, a fundamental barrier for all existing $h$-only methods, we demonstrate that joint $hr$-adaptation demonstrates the ability of an order of magnitude improvement over single adaptivity based methods.
\end{itemize}

These contributions establish that treating mesh adaptation as a unified optimization problem, rather than sequential $h$- and $r$-steps, could open up new possibilities of what is achievable in automated mesh generation.


\section{Mathematical Framework}
\label{sec:math_framework}

This section establishes the mathematical setting for our work. We begin by introducing the abstract stationary PDE problem and its FEM discretization. Then, we provide formal definitions for mesh relocation ($r$-adaptivity) and mesh refinement ($h$-adaptivity), framing them as two distinct optimization problems. Finally, we outline the conventional iterative process that combines these two techniques, setting the stage for the unified approach proposed in this paper.

\subsection{PDE Setting}
\label{sec:abstract_pde_fem}

Let $\Omega \subset \mathbb{R}^d$ (with $d=2$) be a bounded domain, which we refer to as the physical domain. Within this work, we consider PDEs parametrized by a set of parameters, namely $\boldsymbol{\mu} \in \mathcal{P} \subset \mathbb{R}^p$, with $\mathcal{P}$ compact:
$$
    \left\{
    \begin{array}{rll}
        \mathcal{N}[\boldsymbol{\mu}](u) &= 0 \qquad & \mbox{in} \ \Omega, \\
        \mathcal{B}[\boldsymbol{\mu}](u) &= 0  \qquad & \mbox{on} \ \partial{\Omega},
    \end{array}
    \label{eq:strong_form}
    \right. 
$$
where $\mathcal{N}$ is a generic nonlinear differential operator and $\mathcal{B}$ enforces the boundary conditions. 
To solve problem \eqref{eq:strong_form} numerically, we introduce a triangulation
$\mathcal{T}$ of the computational domain and define
$\tilde{\Omega} = \operatorname{int}\left(\bigcup_{K \in \mathcal{T}} K \right)$, with $\tilde{\Omega} \approx \Omega$.
On the mesh $\mathcal{T}$, we consider a finite-dimensional Galerkin space
$X_h$ of dimension $\dim(X_h) = N_h$; here, by $h$ we denote a parameter related to the mesh size of the computational grid.

By applying the Galerkin-FEM on $X_h$, the discretized problem reads as follows:
$$
    \label{eq:general_formulation_fom}
    \mathbf{N}[\boldsymbol{\mu}](\mathbf{u}) = 0 .
$$

We remark that the mesh $\mathcal{T}$ is defined by a set of vertices (or nodes) $\mathcal{Z} = \{\boldsymbol{z}_i\}_{i=1}^{N_v}$, with $N_v = N_h$, and their connectivity matrix $\mathcal{C}$, which defines the elements $\{K\}_{i=1}^{N_e}$. The goal of mesh adaptivity, or adaptive meshing, is to construct a mesh $\mathcal{T}$ by minimizing a suitable error metric $E_{\mathcal{T}} = E(\{u(\mathbf{z}_i)\}_{i=1}^{N_v}, \mathbf{u})$, taking into account the discretization error measured in a suitable norm. 
In this setting, posteriori error estimators are often used to guide the adaptive meshing by providing local error indicators $\{\eta_K\}_{K \in \mathcal{T}}$, which approximate the contribution of each element $K$ to the total error.

\subsection{\(r\)-Adaptivity: Mesh Relocation via Optimal Transport}
\label{sec:r_adaptivity_mapping}

The goal of \(r\)-adaptivity, or mesh relocation, is to adjust the spatial positions of the mesh vertices $\mathcal{Z}$ to improve approximation accuracy, while keeping the number of vertices and the mesh connectivity fixed. This is formally achieved by identifying a mapping from a fixed reference domain to the physical domain.

Let $\Omega_F$ be a computational domain (e.g., the unit square) with a fixed, uniform mesh. Mesh relocation is defined by a differentiable and invertible deformation map $\mathbf{\Phi}: \Omega_F \to \tilde{\Omega}$ which transforms the fixed vertex coordinates $\boldsymbol{\xi}_i \in \Omega_F$ to new vertex coordinates $\boldsymbol{z}_i = \mathbf{\Phi}(\boldsymbol{\xi}_i)$ in the domain $\tilde{\Omega}$. The adapted mesh is then $\mathcal{T}_{\mathbf{\Phi}}$, formed by the new vertices $\mathcal{Z}_{\mathbf{\Phi}} =  \{ \mathbf{z}_i \}_{i=1}^{N_v} = \{\mathbf{\Phi}(\boldsymbol{\xi}_i)\}_{i=1}^{N_v}$ and the original connectivity $\mathcal{C}$. The main idea to construct an effective map $\mathbf{\Phi}$ is to rely on the \textit{equidistribution principle}~\cite{budd2009adaptivity,huang2011adaptive}. This principle aims at uniformly distributing a scalar monitor function $m : \tilde{\Omega} \rightarrow \mathbb{R}$, such that $m(\boldsymbol{z}) > 0 \; \forall\mathbf{z} \in \tilde{\Omega}$,  in the computational domain. The monitor function (such as an apriori estimate of the interpolation error, or the curvature of the solution, through the norm of the hessian) is designed to have large values in regions of the domain where fine mesh resolution, i.e., small elements' size, is needed. The equidistribution principle is expressed as:
$$
m(\mathbf{\Phi}(\boldsymbol{\xi})) \det(J_{\mathbf{\Phi}}(\boldsymbol{\xi})) = \theta,
$$
where $J_{\mathbf{\Phi}}$ is the Jacobian of the map $\mathbf{\Phi}$ and $\theta$ is a normalization constant defined as $\theta = | \Omega_F |^{-1} \int_{\tilde{\Omega}} m(\boldsymbol{z}) d\boldsymbol{z}$.

However, the equidistribution principle sets the desired mesh density but it does not uniquely define the map $\mathbf{\Phi}$ in dimensions $d > 1$. To obtain a unique and well-behaved, i.e., non-tangling mesh, this condition is coupled with an optimal transport requirement~\cite{budd2009adaptivity}. More precisely, the map $\mathbf{\Phi}$ is chosen to minimize the transport cost functional. For the quadratic cost, this corresponds to minimize the squared $L^2(\Omega_F)$ norm of the displacement field
$
I_2 = \int_{\Omega_F} |\mathbf{\Phi}(\boldsymbol{\xi}) - \boldsymbol{\xi}|^2 d\boldsymbol{\xi}.
$ By Brenier's theorem~\cite{brenier1991polar}, the unique optimal transport map for a quadratic cost is the gradient of a convex scalar potential $\phi: \Omega_F \to \mathbb{R}$, such that $\mathbf{\Phi} = \nabla_{\boldsymbol{\xi}} \phi$. Substituting this expression into the equidistribution equation yields the fully nonlinear Monge-Ampère equation for the potential $\phi$:
$$
\det(D^2 \phi(\boldsymbol{\xi})) = \frac{\theta}{m(\nabla_{\boldsymbol{\xi}} \phi(\boldsymbol{\xi}))}.
$$
Solving this elliptic PDE for $\phi$ defines the optimal vertex locations~\cite{budd2009adaptivity}. However, solving it directly is computationally expensive and practical alternatives include relaxation methods, like the Parabolic Monge-Ampère (PMA) formulation, and variational mesh energy minimization~\cite{delzanno2008optimal,huang2011adaptive}.

Ultimately, we note that a \(r\)-adaptivity problem can be expressed as an optimization problem where the goal is to find an optimal valid mapping $\mathbf{\Phi}^*$ minimizing the error:
$$
\mathbf{\Phi}^* = \argmin_{\mathbf{\Phi}:\,\text{valid}} E_{\mathcal{T}},
$$
where a mapping is considered valid if it is an orientation-preserving diffeomorphism. This physically means that elements do not invert or overlap, which is mathematically guaranteed if the determinant of the map's Jacobian, $\det(J_{\mathbf{\Phi}})$, remains positive throughout the domain.

\subsection{\(h\)-Adaptivity: Mesh Refinement as Decision Problem}
\label{sec:h_adaptivity_functional}

\(h\)-adaptivity, or mesh refinement, aims at modifying the mesh topology by selectively subdividing elements to increase resolution locally. It can be formalized as a discrete decision problem.

Let the current mesh be $\mathcal{T} = (\mathcal{Z}, \mathcal{C})$, where $\mathcal{Z}$ is the set of vertices and $\mathcal{C}$ is the connectivity of the elements. We introduce a binary decision vector $\boldsymbol{b} = \{b_K\}_{K \in \mathcal{T}} \in \mathbb{R}^{N_e}$, where $b_K \in \{0, 1\} \textnormal{ such that }b_K = 1$ means element refinement, and $b_K = 0$ no change. Thus, applying the refinement decisions in $\boldsymbol{b}$ produces a new mesh $\mathcal{T}_{\boldsymbol{b}} = (\mathcal{Z}_{\boldsymbol{b}}, \mathcal{C}_{\boldsymbol{b}})$. The new element set $\mathcal{C}_{\boldsymbol{b}}$ consists of the original unmarked elements and the children of the marked elements, while the new vertex set $\mathcal{Z}_{\boldsymbol{b}}$ includes new vertices created during subdivision. Conformity is maintained by ensuring no hanging nodes, which in 2D is handled by conformity-preserving procedures like red-green-blue refinement that refines adjacent elements as needed~\cite{bank1983some}.

The optimal refinement strategy seeks to balance error reduction with computational cost. This can be formulated as the minimization of an objective functional:
$$
\mathcal{J}(\boldsymbol{b}) = E_{\mathcal{T}_{\boldsymbol{b}}} + \alpha \, \text{Cost}(\boldsymbol{b}),
$$
where $\text{Cost}(\boldsymbol{b})$ is a penalty term for the increase in the number of DOFs, and $\alpha > 0$ is a weighting parameter. The ideal set of refinements, for one iteration, is given by the solution to the combinatorial optimization problem:
$$
\boldsymbol{b}^* = \argmin_{\boldsymbol{b} \in \{0,1\}^{N_v}} \mathcal{J}(\boldsymbol{b}).
$$
Classical numerical algorithms, such as those using Dörfler marking~\cite{dorfler1996convergent}, approximate the solution of the optimization problem by marking all elements whose local error indicators $\{\eta_K\}_{K \in \mathcal{T}}$ contribute to a fixed fraction of the total error resulting in a short-term and greedy approach. In contrast, a learning-based approach can train a policy $\pi$ to predict the decision $b_K$ for each element, aiming to directly approximate $\boldsymbol{b}^*$ by learning the long-term trade-off between error and cost.

\subsection{\(hr\)-Adaptivity}
\label{sec:combined_hr}

Combining \(r\)- and \(h\)-adaptivity results in a \(hr\)-adaptive strategy, which is typically performed as an iterative loop. In particular, given a mesh $\mathcal{T}^{(n)}$ at iteration $n$, the process is as follows:

\begin{enumerate}
    \item {Solve:} Compute the discrete solution $\mathbf{u}_{\mathcal{T}}^{(n)}$ on the current mesh $\mathcal{T}^{(n)}$.
    \item {\(r\)-Step (Relocation):} Determine a valid relocation map $\mathbf{\Phi}^{(n)}$ based on $\mathbf{u}_{\mathcal{T}}^{(n)}$ to produce an intermediate moved mesh $\widetilde{\mathcal{T}}^{(n)} = \mathcal{T}_{\mathbf{\Phi}^{(n)}}$.
    \item {(Optional) Solve on Moved Mesh:} Resolve or estimate the approximation error on $\widetilde{\mathcal{T}}^{(n)}$ to obtain updated error indicators for the refinement step.
    \item {\(h\)-Step (Refinement):} Using error indicators from $\widetilde{\mathcal{T}}^{(n)}$ (or $\mathcal{T}^{(n)}$), determine a refinement vector $\boldsymbol{b}^{(n)}$ to generate the mesh for the next iteration, $\mathcal{T}^{(n+1)} = (\widetilde{\mathcal{T}}^{(n)})_{\boldsymbol{b}^{(n)}}$.
    \item {Update:} Set $n \leftarrow n+1$ and repeat until a stopping criterion is satisfied (e.g., error tolerance or computational budget).
\end{enumerate}

In a fully unified framework, one could seek to find the pair $(\mathbf{\Phi}, \boldsymbol{b})$ that jointly optimizes a combined objective:
\begin{equation}
\mathcal{J}(\mathbf{\Phi}, \boldsymbol{b}) = E_{\mathcal{T}_{\mathbf{\Phi},\boldsymbol{b}}} + \alpha \, \text{Cost}(\boldsymbol{b}),
\label{eq:cost_functional}
\end{equation}
where $\mathcal{T}_{\mathbf{\Phi},\boldsymbol{b}}$ is the mesh created by first applying the map $\mathbf{\Phi}$ and then refining according to $\boldsymbol{b}$. Since solving this combined problem is intractable, practical methods decouple the two adaptive steps. However, the decisions can be strongly coupled, for instance, through a shared learned policy that considers both vertex- and element-based features to determine mesh movement and refinement simultaneously. Our proposed work explores precisely this argument by casting the \(hr\)-adaptivity problem in a deep RL framework, where a joint policy over $(\mathbf{\Phi}, \boldsymbol{b})$ is learned to optimize the long-term error-cost trade-off.


\section{Unified Learning Framework for \(hr\)-Adaptivity}
\label{sec:methodology}

In this section, we introduce \textit{HypeR}, a deep RL framework designed to solve the combined \(hr\)-adaptivity problem defined in Section \ref{sec:math_framework}. HypeR learns a single, unified policy that jointly determines the vertex relocation map $\mathbf{\Phi}$ and the element refinement vector $\boldsymbol{b}$ in a single iterative loop. This approach is in contrast with classical numerical methods that treat relocation and refinement as separate, sequential steps.

\subsection{Mesh Representation as a Hypergraph}
\label{sec:mesh_rep}

Let $\mathcal{T}^{(n)}$ be a conforming mesh of the domain $\tilde{\Omega} \subset \mathbb{R}^d$ at the $n$-th step of the adaptive mesh refinement process. Here, $n$ indexes the successive iterations of our $hr$-adaptive algorithm, where each iteration involves: (1) solving the PDE on the current mesh, (2) applying the learned relocation policy to move vertices, and (3) applying the learned refinement policy to subdivide selected elements. The mesh is defined by its set of vertices $\mathcal{Z}^{(n)} = \{\boldsymbol{z}_i^{(n)}\}_{i=1}^{N_v^{(n)}}$ and its set of elements (e.g., triangles or tetrahedra) $\{K_j^{(n)}\}_{j=1}^{N_e^{(n)}}$\footnote{The number of vertices $N_v^{(n)}$ and elements $N_e^{(n)}$ change with each adaptation step $n$. For the sake of clarity, we hereafter omit the superscript $(n)$ from $N_v$ and $N_e$.}. To represent the mesh in a form suitable for processing by GNNs and hypergraph neural networks, we encode $\mathcal{T}^{(n)}$ as a hypergraph $\mathcal{H}^{(n)}$, as illustrated in Figure~\ref{fig:hypergraph_representation}. This encoding creates a one-to-one mapping where each mesh vertex becomes a hypergraph node and each mesh element a hyperedge connecting its defining  nodes. Formally, this structure is defined as:

\begin{figure}[t]
  \centering
  \includegraphics[width=1\linewidth]{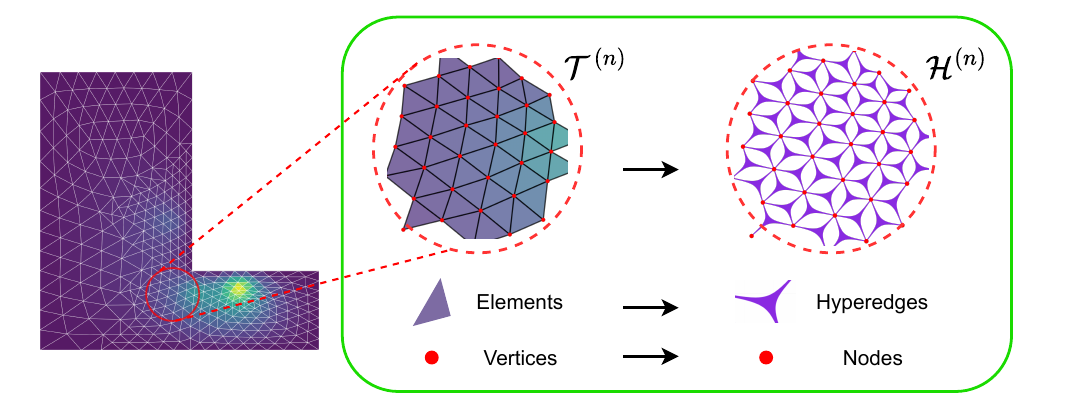}
\caption{Hypergraph representation of the mesh $\mathcal{T}^{(n)}$. A zoomed-in region of the mesh on the L-shaped domain shows its modeling by the hypergraph $\mathcal{H}^{(n)}$. This encoding creates a one-to-one mapping where each mesh vertex (red dot) is treated as a hypergraph node, and each triangular element is represented as a hypergraph hyperedge (purple) connecting its constituent nodes.}
 \label{fig:hypergraph_representation}
\end{figure}

$$
 \mathcal{H}^{(n)} = \bigl(\mathcal{Z}^{(n)}, \{K_j^{(n)}\}_j, H^{(n)}, A^{(n)}\bigr),
$$
where $H^{(n)} \in \{0,1\}^{N_v \times N_e}$ is the incidence matrix with $[H^{(n)}]_{ij} = 1$ if vertex $\boldsymbol{z}_i^{(n)}$ belongs to element $K_j^{(n)}$, and $A^{(n)} \in \{0,1\}^{N_v \times N_v}$ is the adjacency matrix where $[A^{(n)}]_{ik} = 1$ if vertices $\boldsymbol{z}_i^{(n)}$ and $\boldsymbol{z}_k^{(n)}$ share at least one element. These matrices encode the complete mesh topology.

To capture the geometrical features of the mesh and the physical state of the PDE solution, each vertex $\boldsymbol{z}_i^{(n)}$ and element $K_j^{(n)}$ are augmented with additional features as follows:
$$
 \boldsymbol{f}_{\boldsymbol{z}_i}^{(n)} = \bigl[\boldsymbol{z}_i^{(n)}, \mathbf{u}^{(n)}_i \bigr] \in \mathbb R^{d_v}, \quad \boldsymbol{f}_{K_j}^{(n)} = \bigl[|K_j^{(n)}|, \text{shape metrics}, \text{solution variability metrics}\bigr] \in \mathbb R^{d_e}
$$
where $\mathbf{u}^{(n)}_i$ is the FEM solution at vertex $\boldsymbol{z}_i^{(n)}$, and $|K_j^{(n)}|$ is the area (or volume) of element $K_j^{(n)}$. The element shape and solution variability metrics are chosen to quantify mesh quality and local solution behavior, respectively (see Appendix~\ref{app_sec:mesh_features} for more details).

In this way, the hypergraph representation $\mathcal{H}^{(n)}$ provides a unified data structure that encapsulates the mesh topology (via $H^{(n)}, A^{(n)}$), its geometry, including vertex coordinates, element areas and skewness, and the physical solution field (via $\boldsymbol{f}_{\boldsymbol{z}_i}^{(n)}, \boldsymbol{f}_{K}^{(n)} $), preparing it for processing by the GNN-based policy network. It is crucial to note that this representation is dynamic: as $h$-adaptivity subdivides elements, new vertices are created and parent elements are replaced by their children. Consequently, the dimensions of the incidence matrix $H^{(n)}$ and adjacency matrix $A^{(n)}$ must expand, by adding new rows for the vertices and updating the column structure to reflect the new elements. This changing agent population necessitates a formal mechanism to trace lineage and map rewards across steps, a challenge addressed in the following section.

\subsection{Adaptive Dual-Swarm Markov Decision Process}
\label{sec:mdp}

We frame the problem of finding an optimal sequence of \(hr\)-adapted meshes as a MDP. The objective is to learn a policy $\pi_\theta$ that, given the current mesh, selects a near-optimal pair $(\mathbf{\Phi}^{(n)}, \boldsymbol{b}^{(n)})$ to minimize the long-term, cumulative objective functional $\mathcal{J}$ introduced in Section \ref{sec:combined_hr} in \eqref{eq:cost_functional}. This is achieved by treating each mesh vertex and element as an independent actor-critic in a swarm of agents. In this paradigm, each agent's "actor" component learns a policy to select an action (a vertex move or an element split), while its "critic" component learns to estimate the long-term value of taking that action from the current state.

Formally, the state at iteration $n$ is the mesh hypergraph $S^{(n)} = \mathcal{H}^{(n)}$. The action $A^{(n)} = (A^{(n), r}, A^{(n), h})$ is a joint decision composed of contributions from the two swarms. The refinement action $A^{(n), h}$ is the binary decision vector $\boldsymbol{b}^{(n)} = \{a_{K_j}^{(n)}\}_{j=1}^{N_e}$. The relocation action $A^{(n), r}$ is realized through the vertex policy $\pi_\theta^r$, which learns to approximate the optimal map $\mathbf{\Phi}^*$ introduced in Section \ref{sec:r_adaptivity_mapping}. At each step $n$, the vertex policy takes the current state $S^{(n)}$ and outputs a distribution over new coordinates for each vertex. The action for a single vertex, $\boldsymbol{a}_{\boldsymbol{z}_i}^{(n)} \in \mathbb{R}^d$, is a sample from this distribution representing its proposed new location. The collective set of actions $A^{(n), r} = \{\boldsymbol{a}_{\boldsymbol{z}_i}^{(n)}\}_{i=1}^{N_v}$ thus defines the relocated vertex set $\mathcal{Z}^{(n+1)}$ for the subsequent mesh.

A central challenge in the definition of such an MDP is to keep track of the agent hierarchy as the elements are refined and new vertices are added. To assign credit correctly, we define projection maps that trace an agent's lineage. For an element $K_j^{(n)}$ at iteration time $n$, we use a map $\phi^h_{n\to n+k}(j,j')$ that traces its descendants $\{K_{j'}\}$ at time $n+k$, following the same procedure as in ASMR++ \citep[Sec.~3.1]{freymuth2024asmrplus}. However, for vertex agents a similar "complete" projection is computationally prohibitive. Since $h$-refinement creates new vertices at each step, a full credit assignment would necessitate a complex, multi-step reward aggregation backward through time. If we consider a 2D triangle refined element, it not only generates 4 new elements but also adds 3 new vertices, shared with neighboring elements. So, a reward received by a new vertex created at step $n+k-1$, realized at $n+k$, would first need to be distributed back to all its parent vertices at $n+k-1$. This aggregation process would then need to be recursively applied for every preceding step ($n+k-2, \dots, n$) to correctly route all future reward signals back to the original agents. Tracking this dynamic, recursive credit flow for every agent across every trajectory is infeasible in a large-scale learning setting. To overcome this issue, we adopt a practical approximation motivated by the short-term nature of the relocation task. Unlike element refinement, which permanently alters the mesh topology with long-term consequences, vertex relocation ($r$-adaptivity) is a local tuning of the current geometry whose benefit is realized entirely in the immediate error reduction of the next step. Instead of a full projection, we use a simple identity mapping for vertices:
$$
 \phi^r_{n\to n+k}(i,i') = \begin{cases} 1, & \text{if } \boldsymbol{z}_{i'}^{(n+k)} \text{ and } \boldsymbol{z}_i^{(n)} \text{ refer to the same persistent vertex}, \\ 0, & \text{otherwise}. \end{cases}
$$

This map is "incomplete" because it only attributes rewards from vertices that persist from iteration $n$, effectively ignoring the reward contributions from any vertices created in subsequent refinement steps. However, since the primary goal of vertex relocation is immediate, local error reduction, we can pair this simplified map with a very small discount factor $\gamma_r \ll 1$. This enforces a short-term learning objective where the discounted value of future rewards - and thus the error introduced by the incomplete projection - becomes negligible. Therefore, the vertex agents' discounted long term reward reduces too:
$$
 G^{(n), r}_{\boldsymbol{z}_i} = \sum_{k=0}^\infty \gamma_r^k \sum_{\boldsymbol{z}_{i'} \in \mathcal{Z}^{(n+k)}} \phi^r_{n\to n+k}(i,i')\,r^{(n+k), r}_{\boldsymbol{z}_{i'}} \approx r^{(n), r}_{\boldsymbol{z}_i}.
$$
In contrast, the element agents (\(h\)-adaptivity) use a large discount factor $\gamma_h \approx 1$ and the full projection map to capture the long-term consequences of refinement:
$$
 G^{(n), h}_{K_j} = \sum_{k=0}^\infty \gamma_h^k \sum_{K_{j'} \in \mathcal{T}^{(n+k)}} \phi^h_{n\to n+k}(j,j')\,r^{(n+k), h}_{K_{j'}}.
$$
Then, the policy is factorized across the two swarms, assuming conditional independence given the global mesh state $\mathcal{H}^{(n)}$:
$$
 \pi_\theta(A^{(n)}|S^{(n)}) = \prod_{\boldsymbol{z}_i \in \mathcal{Z}^{(n)}} \pi_\theta^r(\boldsymbol{a}^{(n)}_{\boldsymbol{z}_i} | \mathcal{H}^{(n)}) \times \prod_{K_j \in \mathcal{T}^{(n)}} \pi_\theta^h(a^{(n)}_{K_j} | \mathcal{H}^{(n)}).
 \label{eq:policy}
$$
The policy in \eqref{eq:policy} is optimized via a multi-objective PPO  algorithm~\cite{schulman2017proximal} to simultaneously maximize the expected returns $\mathbb{E}[G^r]$ and $\mathbb{E}[G^h]$.


\subsection{Neural Network Architecture}
\label{sec:arch}

\begin{figure}[t]
  \centering
  \includegraphics[width=1\linewidth]{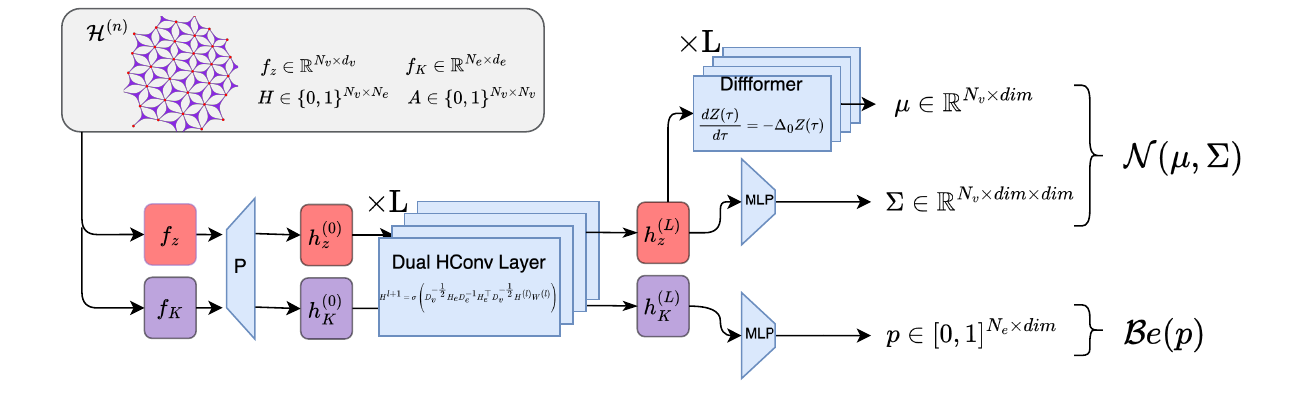}
\caption{HypeR policy (actor) network architecture. The mesh hypergraph $\mathcal{H}^{(n)}$ provides vertex features ($f_z$) and element features ($f_K$), which are first projected onto a latent space by a linear embedding layer (P). The resulting embeddings ($h_z^{(0)}, h_K^{(0)}$) are processed by $L$ dual hypergraph convolutional (HConv) layers to produce the final embeddings $h_z^{(L)}$ and $h_K^{(L)}$. These are fed into two distinct agent policy heads: (1) the vertex-agent head generates a multivariate Gaussian policy $\mathcal{N}(\mu, \Sigma)$ by using a stack of $L$ Difformer layers to produce the mean $\mu$ and a separate MLP to produce the covariance $\Sigma$, and (2) the element-agent head uses a MLP to generate a probability vector $p$ for the Bernoulli refinement policy $\mathcal{B}e(p)$. The corresponding value (critic) heads, which share the same hypergraph backbone, are omitted for clarity.}
 \label{fig:network_architecture}
\end{figure}

The joint policy $\pi_{\theta}(A^{(n)}|S^{(n)})$ and its associated value functions are parameterized by a single actor-critic neural network.The architecture of the policy (actor) component is illustrated in Figure~\ref{fig:network_architecture}. This network is designed to process the full mesh hypergraph $\mathcal{H}^{(n)}$, leveraging a shared hypergraph-convolutional backbone to extract rich features before feeding them into specialized output heads for each agent swarm (vertices and elements). This architecture allows the network to learn complex dependencies between vertex positions, element shapes, and the local behavior of the PDE solution. The architectural components are detailed below.

\paragraph{Initial feature transformation.}
The input feature vectors for vertices, $\boldsymbol{f}_{\boldsymbol{z}_i}^{(n)} \in \mathbb{R}^{d_v}$, and elements, $\boldsymbol{f}_{K_j}^{(n)} \in \mathbb{R}^{d_e}$\footnote{Throughout this section, we suppress the step superscript $^{(n)}$ from vertex and element notation for readability. All quantities $\boldsymbol{z}_i$, $K_j$, and their associated embeddings are implicitly at adaptation step $n$.}, are first projected into a $D$-dimensional latent space using separate linear transformations. This creates the initial embeddings for the subsequent message passing layers
$$
 \boldsymbol{h}_{\boldsymbol{z}_i}^{(0)} = W_v \boldsymbol{f}_{\boldsymbol{z}_i}^{(n)} \quad \textnormal{and} \quad 
 \boldsymbol{h}_{K_j}^{(0)} = W_e \boldsymbol{f}_{K_j}^{(n)},
$$
where $W_v \in \mathbb{R}^{D \times d_v}$ and $W_e \in \mathbb{R}^{D \times d_e}$ are learnable weight matrices.

\paragraph{Hypergraph convolutional operations.}
The core of the neural network consists of $L$ layers of dual hypergraph convolution (HConv) (see Appendix~\ref{app_sec:hconv} for more details), which enables message passing between vertices and elements. At each layer $\ell \in \{1, \dots, L\}$, embeddings are updated via a two-stage process: \textit{Vertex-to-Element}, where information is passed from vertices to their incident elements to update the element embeddings, and \textit{Element-to-Vertex}, where information is then passed back from the newly updated element embeddings to their incident vertices
$$
    \widetilde{\boldsymbol{h}}_{K_j}^{(\ell)} = \text{HConv}_{\mathcal{Z} \to \mathcal{K}}\bigl(\{\boldsymbol{h}_{\boldsymbol{z}_i}^{(\ell-1)}\}  \{\boldsymbol{h}_{K_j}^{(\ell-1)}\}\bigr)
     \quad \textnormal{and} \quad 
    \widetilde{\boldsymbol{h}}_{\boldsymbol{z}_i}^{(\ell)} = \text{HConv}_{\mathcal{K} \to \mathcal{Z}}\bigl(\{\boldsymbol{h}_{\boldsymbol{z}_i}^{(\ell-1)}\}, \{\widetilde{\boldsymbol{h}}_{K_j}^{(\ell)}\}\bigr).
$$

Each message passing step is followed by a point-wise nonlinearity (i.e., ReLU) to produce the final embeddings for that layer, $\boldsymbol{h}_{\boldsymbol{z}_i}^{(\ell)}$ and $\boldsymbol{h}_{K_j}^{(\ell)}$. After $L$ layers, the final embeddings $\boldsymbol{h}_{\boldsymbol{z}_i}^{(L)}$ and $\boldsymbol{h}_{K_j}^{(L)}$ encode contextual information about the $L-$hop neighborhood of each entity within the mesh.

\paragraph{Output heads.}
The final embeddings are processed by separate heads for each agent type to produce action distributions and value estimates.

\smallskip\noindent
{\underline{Element agent heads.}} For each element $K_j$, its final embedding $\boldsymbol{h}_{K_j}^{(L)}$ is passed to:
\begin{itemize}
    \item {Policy head:} A two-layer MLP, i.e., $\text{MLP}_{\text{elem}}^{\pi}$, that outputs a logit $\ell_{K_j} \in \mathbb{R}$. This logit parameterizes a Bernoulli distribution for the binary refinement action $a_{K_j} \in \{0, 1\}$, which consists of a component of the refinement vector $\boldsymbol{b}^{(n)}$;
    \item {Value head:} A separate two-layer MLP, i.e., $\text{MLP}_{\text{elem}}^{V}$, that provides as output a scalar value estimate $V^e_{K_j} \in \mathbb{R}$.
\end{itemize}

\smallskip\noindent
\underline{Vertex agent heads.} For each vertex $\boldsymbol{z}_i$, its final embedding $\boldsymbol{h}_{\boldsymbol{z}_i}^{(L)}$ is passed to:
\begin{itemize}
    \item {Policy head:} This head determines the parameters definying a multivariate Gaussian distribution over the vertex's new coordinates in $\mathbb{R}^d$. It consists of two main components. The first one is a \textit{diffformer block}, first introduced in G-Adaptivity~\cite{rowbottom2025gadaptivity}, which takes the vertex's current coordinates $\boldsymbol{z}_i$, together with its embedding $\boldsymbol{h}_{\boldsymbol{z}_i}^{(L)}$, and provides as output the mean of the distribution $\boldsymbol{\mu}_i \in \mathbb{R}^d$
    \footnote{More details on the difformer block can be found in Section \ref{sec:diffformer}.}.
    The second component consists of a separate MLP which produces the covariance matrix $\boldsymbol{\Sigma}_i$. Thus, the policy for the vertex agent is  $\pi_\theta^r(\cdot | \mathcal{H}^{(n)}) = \mathcal{N}(\boldsymbol{\mu}_i, \boldsymbol{\Sigma}_i)$;
    \item {Value head:} A two-layer MLP, i.e., $\text{MLP}_{\text{vert}}^{V}$, that gives as output a scalar value estimate $V^v_{\boldsymbol{z}_i} \in \mathbb{R}$.
\end{itemize}

In summary, the overall network is fed with the hypergraph $\mathcal{H}^{(n)}$ and outputs, for each vertex, the parameters of a Gaussian distribution $\mathcal{N}(\boldsymbol{\mu}_i, \boldsymbol{\Sigma}_i)$ to guide relocation, and for each element, a Bernoulli probability to guide refinement. This process directly translates the high-level representation of the state of the mesh to the low-level actions that define the \(hr\)-adaptation step.


\subsubsection{Diffformer Block and Non-Tangling Guarantee}
\label{sec:diffformer}

The vertex relocation map $\mathbf{\Phi}^{(n)}$ is sampled from the multivariate Gaussian distribution output of the vertex agent policy head. This is built from the \textit{diffformer} operator, a GNN block, first introduced in G-Adaptivity~\cite{rowbottom2025gadaptivity}, inspired by classical velocity-based methods and in which vertex relocation is casted as a graph diffusion process. More precisely, the evolution of vertex coordinates $\mathcal{Z}(\tau)$ over a continuous pseudo-time $\tau$ is governed by the ordinary differential equation (ODE):
$$
 \frac{d\mathcal{Z}(\tau)}{d\tau} = (A_{\theta} - I)\mathcal{Z}(\tau) = -\Delta_{\theta}\mathcal{Z}(\tau),
$$
initialized with the current mesh coordinates, i.e., $\mathcal{Z}(0) = \mathcal{Z}^{(n)}$. Here, $\Delta_{\theta}$ is a learnable, weighted graph Laplacian (see \citet{rowbottom2025gadaptivity} for full details). The row-stochastic attention matrix $A_{\theta}$ is computed dynamically from the final vertex embeddings $\{\boldsymbol{h}_{\boldsymbol{z}_i}^{(L)}\}_{i=1}^{N_v}$ provided by the shared hypergraph backbone. This allows the diffusion to be anisotropic and data-driven, conditioned on the rich state information captured by the hypergraph.

In practice, this ODE is integrated over a fixed interval using a forward Euler scheme with step size $\delta\tau$. The key result, established in G-Adaptivity~\cite{rowbottom2025gadaptivity}, is that the discrete evolution is guaranteed to prevent the critical failure mode of mesh tangling, in which elements become inverted and geometrically invalid. For a sufficiently small step size ($\delta\tau < 0.5$), the transformation matrix $(I - \delta\tau\Delta_\theta)$ has a positive determinant, which ensures that all element orientations are preserved throughout the relocation. Intuitively, the diffusion process moves each vertex into the convex hull of its neighbors, making element inversion impossible.

Within HypeR, the final vertex positions $\mathcal{Z}(T)$ computed by the diffformer serve as the mean $\boldsymbol{\mu}$ for the Gaussian vertex policy. 
During the training phase, agent actions sampled from the vertex agent policy head that cause mesh tangling are penalized (see Appendix \ref{app_sec:implementation}). At inference, we take the mode of this distribution, which is the direct output of the difformer. The relocation map $\mathbf{\Phi}^{(n)}$ thus inherits this non-tangling property, producing guaranteed valid meshes.


\subsection{Reinforcement Learning Procedure and Mesh Update}
\label{sec:rl_mesh_update}

\begin{figure}[t]
  \centering
  \includegraphics[width=1\linewidth]{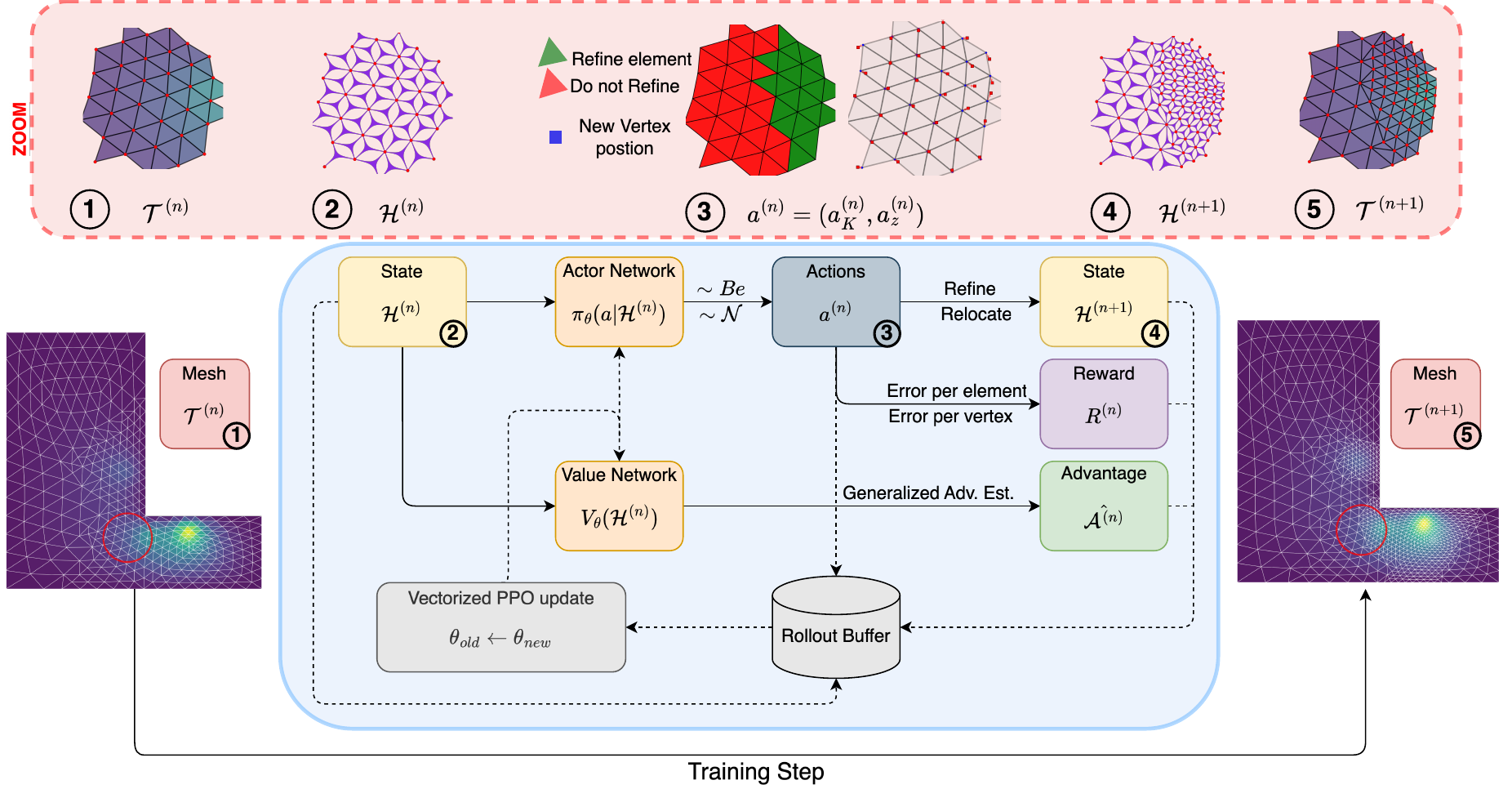}
\caption{Overview of one HypeR training step, transitioning from mesh $\mathcal{T}^{(n)}$ (1) to $\mathcal{T}^{(n+1)}$ (5). The PPO loop (blue box) begins with the hypergraph state $\mathcal{H}^{(n)}$ (2), which is fed into the joint actor-critic network. The network outputs policy distributions and value estimates $V^{(n)}$. The policy $\pi_{\theta}$ samples a joint action $A^{(n)}$ (3), consisting of continuous vertex relocations ($\sim \mathcal{N}$) and binary element refinements ($\sim \text{Bernoulli}$). Applying these actions produces the intermediate relocated mesh $\widetilde{\mathcal{T}}^{(n)}$ (4) and the final refined mesh $\mathcal{T}^{(n+1)}$ (5). Per-agent rewards $R^{(n)}$ are computed via FEM solves. The full transition tuple (state, action, reward, value) is stored in the rollout buffer. Once the trajectory is complete, GAE is applied retroactively to compute the advantages $\hat{\mathcal{A}}^{(n)}$ used for the vectorized PPO update (dashed line).}
 \label{fig:rl_loop}
\end{figure}

In the following, we describe the data collection (rollout) process and the policy update step. The entire procedure for a single training step is illustrated in Figure~\ref{fig:rl_loop}. This diagram shows how the agent transitions the mesh from a state $\mathcal{T}^{(n)}$ to the next state $\mathcal{T}^{(n+1)}$ by applying the learned policy, and how the resulting information is used to update the network via PPO.

\paragraph{Rollout structure.}
Training data is collected by executing the policy in the simulation environment for trajectories of a fixed length $N$. A single trajectory consists of $N$ sequential mesh adaptation steps. Within each trajectory, steps $n=0, \dots, N-2$ involve a full \(hr\)-adaptation (relocation followed by refinement). The final step, $n=N-1$, performs only the relocation step.

\paragraph{One \(hr\)-adaptivity step.}
A single, complete adaptive step from mesh $\mathcal{T}^{(n)}$ to $\mathcal{T}^{(n+1)}$ unfolds as follows:

\begin{enumerate}
    \item {State Construction:} Given the current mesh $\mathcal{T}^{(n)}$, the governing PDE is solved to obtain the discrete solution $u_{\mathcal{T}^{(n)}}$. From this solution, we compute per-vertex and per-element features (e.g., approximation errors, shape metrics), which are assembled into the hypergraph state representation $S^{(n)} = \mathcal{H}^{(n)}$.

    \item {Network Forward Pass (Policies \& Values):} The state $S^{(n)}$ is passed through the actor-critic network described in \ref{sec:arch}. The architecture produces four distinct outputs required for the PPO algorithm:
    \begin{itemize}[-]
        \item \textit{Vertex Actor:} Parameters for a Gaussian distribution $\mathcal{N}(\boldsymbol{\mu}_i, \boldsymbol{\Sigma}_i)$ for every vertex $\boldsymbol{z}_i$.
        \item \textit{Element Actor:} A Bernoulli probability $p_j = \sigma(\ell_{K_j})$ for every element $K_j$.
        \item \textit{Critics:} Scalar value estimates $V^{(n), r}_{\boldsymbol{z}_i}$ and $V^{(n), h}_{K_j}$ estimating the expected return for vertex and element agents, respectively.
    \end{itemize}

    \item {Joint Action Sampling:} We sample the joint action $A^{(n)} = (A^{(n), r}, A^{(n), h})$ from the distributions parameterized in the previous step.
    \begin{itemize}[-]
        \item \textit{Relocation Action:} New coordinates $\boldsymbol{z}'_i \sim \mathcal{N}(\boldsymbol{\mu}_i, \boldsymbol{\Sigma}_i)$ are sampled to form $A^{(n), r}$.
        \item \textit{Refinement Action:} Binary flags $a_{K_j} \sim \text{Bernoulli}(p_j)$ are sampled to form the refinement vector $\boldsymbol{b}^{(n)} = A^{(n), h}$.
    \end{itemize}

    \item {Mesh Relocation \& Intermediate Reward:} The relocation action $A^{(n), r}$ defines the map $\mathbf{\Phi}^{(n)}$, transforming the current mesh into the intermediate mesh $\widetilde{\mathcal{T}}^{(n)} = \mathcal{T}_{\mathbf{\Phi}^{(n)}}$. As described in Sec.~\ref{sec:arch}, boundary constraints are strictly enforced here. To compute the vertex rewards $r_{\boldsymbol{z}_i}^{(n)}$, an intermediate FEM solve is performed on $\widetilde{\mathcal{T}}^{(n)}$ to measure the immediate error reduction. (Note: This intermediate solve is strictly for training reward computation and is omitted during inference).

    \item {Mesh Refinement \& Final Reward:} The refinement vector $\boldsymbol{b}^{(n)}$ is applied to the moved mesh $\widetilde{\mathcal{T}}^{(n)}$. We utilize standard conformity-preserving procedures—red-green-blue refinement in 2D~\cite{carstensen2004adaptive}—to handle hanging nodes, producing the final mesh $\mathcal{T}^{(n+1)}$. A final FEM solve on $\mathcal{T}^{(n+1)}$ yields the element rewards $r_{K_j}^{(n)}$.

    \item {Transition Storage:} The transition is completed by storing the tuple $(\mathcal{H}^{(n)}, A^{(n)}, R^{(n)}, V^{(n)}, \mathcal{H}^{(n+1)})$ in the rollout buffer. Advantage estimation is deferred until the full trajectory is collected; we then compute the advantages $\hat{\mathcal{A}}^{(n)}$ (which quantify how much better a specific action was compared to the state's expected value) using Generalized Advantage Estimation (GAE)~\cite{schulman2015high} by processing the stored sequence of rewards and value estimates in reverse order.
    
\end{enumerate}

\paragraph{Policy update.}
After collecting a batch of trajectories, the network parameters $\theta$ are updated. The goal of this update is to adjust the policy $\pi_{\theta}$ to maximize the expected future rewards (returns) collected by the agents. In RL, this is achieved by minimizing a specialized loss function (or "surrogate objective") that is distinct from the error or cost functions defined in Section \ref{sec:math_framework}. We use the PPO algorithm~\cite{schulman2017proximal}, which defines a composite loss $\mathcal{L}(\theta)$ designed to stably guide the policy toward higher rewards. We optimize a loss function that addresses the multi-objective nature of the dual-swarm problem:
$$
 \mathcal{L}(\theta) = \bigl(L_r^{\pi}(\theta) + c_r L_r^{V}(\theta)\bigr) + \bigl(L_h^{\pi}(\theta) + c_h L_h^{V}(\theta)\bigr) - \alpha_{\text{ent}} H(\pi_\theta),
$$
where $L^{\pi}$ is the clipped surrogate policy loss and $L^{V}$ is the clipped value function loss. The subscripts $r$ and $h$ denote the losses for the relocation (vertex) and refinement (element) agents, respectively. $H(\pi_\theta)$ is an entropy bonus to encourage exploration, and $(c_r, c_h, \alpha_{\text{ent}})$ are scalar weighting coefficients. The gradients of this combined loss are computed and used to update all shared and head-specific network parameters, thus completing one training iteration.

\subsection{Reward Design and Training Curriculum}
\label{sec:reward}

To guide the policy towards minimizing the global objective $\mathcal{J}(\mathbf{\Phi}, \boldsymbol{b})$, defined in abstract form in equation \eqref{eq:cost_functional}, we design separate reward functions for the vertex and element agent swarms. In this respect, the minimization of the  objective $\mathcal{J}$ is actually performed by maximizing the expected total global return $J^{(0)}\approx J^{(0), r} + J^{(0), h}$ of the dual-swarm system. This total return decomposes into the aggregate contributions of the relocation and refinement swarms. These components are defined as the sum of discounted rewards over adaptation steps $n$: $J^{(0), r} = \sum_{n} \gamma_r^n \sum_{\boldsymbol{z}_i} r_{\boldsymbol{z}_i}^{(n), r}$ and $J^{(0), h} = \sum_{n} \gamma_h^n \sum_{K_j} r_{K_j}^{(n), h}$. Consequently, by designing the immediate rewards $r^{(n), \cdot}$ to scale with local error reduction, maximizing $J^{(0)}$ is equivalent to minimizing the cost functional $\mathcal{J}$.

To compute these rewards, we define two distinct error indicators for an element $K \in \mathcal{T}$ by projecting the high-fidelity solution from the reference mesh $\mathcal{T}_{\text{ref}}$. Let $P_K = \{ p_{K^*} : K^* \in \mathcal{T}_{\text{ref}}, K^* \subset K \}$ be the set of centroids of reference elements contained within $K$. For refinement decisions, we employ the discrete $L_\infty$ error $\eta_{K, \infty} = \max_{p \in P_K} |u_{\mathcal{T}}(p) - u_{\text{ref}}(p)|$, consistent with~\cite{freymuth2024asmrplus}. For relocation and global monitoring, we utilize the discrete volume-weighted squared $L_2$ error, approximated as $\eta_{K, 2}^2 = \sum_{p \in P_K} |K^*_p| \, |u_{\mathcal{T}}(p) - u_{\text{ref}}(p)|^2$, where $|K^*_p|$ is the volume of the reference element corresponding to centroid $p$.

These rewards serve as local proxies for error reduction and mesh quality. Moreover, we employ a two-phase training curriculum strategy: \textit{Phase I} trains relocation with random refinement to learn stable geometry-aware features; \textit{Phase II} jointly trains relocation and refinement to enable coordinated policies.

\paragraph{Per-vertex reward for relocation.} The reward for a vertex agent is based on the immediate reduction in local error achieved through the relocation map $\mathbf{\Phi}^{(n)}$. We first define a vertex-centric error for each $\boldsymbol{z}_i \in \mathcal{Z}^{(n)}$ as a volume-weighted average of the dicrete squared $L_2$ error indicators $\eta_{K, 2}^2$ of its incident elements denoted as $\{K \in E(\boldsymbol{z}_i)\}$:
$$
 \varepsilon_i^{(n)} = \frac{\sum_{K \in E(\boldsymbol{z}_i)} |K|\,{\eta_{K, 2}^2}}{\sum_{K \in E(\boldsymbol{z}_i)} |K|}.
$$
After applying the relocation to obtain the moved mesh $\widetilde{\mathcal{T}}^{(n)}$, we recompute this error as $\varepsilon_i^{(n),+}$ and find the difference vector $\boldsymbol{\Delta}$ with components $\Delta_i = \varepsilon_i^{(n)} - \varepsilon_i^{(n),+}$. Since the benefit of moving one vertex effects only its 1-hop neighborhood, we diffuse this raw signal across the vertex adjacency graph $A = A^{(n)}$ for 20 iterations of a Personalized PageRank update~\cite{gasteiger2019diffusion}, with teleport probability $\beta=0.5$, acting as a low pass filter:
$$
 \boldsymbol{r}^{\text{loc}} \leftarrow (1-\beta)\boldsymbol{\Delta} + \beta D^{-1}A\,\boldsymbol{r}^{\text{loc}}.
$$
where $D$ is the diagonal degree matrix. The final vertex reward vector $\boldsymbol{r}^{(n), r}$ is then obtained by normalizing this smoothed signal across the entire vertex set, i.e., $\boldsymbol{r}^{(n), r} = \boldsymbol{r}^{\text{loc}} / \|\boldsymbol{r}^{\text{loc}}\|_{l_\infty}$, to restrict values to $[-1, 1]$. Moreover, if any vertex movement results in a tangled element, the implicated vertices receive a strong penalty of $-1.5$, while all others receive zero.

\paragraph{Per-element reward for refinement.} The reward for an element agent, which guides the selection of the refinement vector $\boldsymbol{b}^{(n)}$, adapts the maximum-error-drop signal from ASMR$^{++}$ \cite{freymuth2024asmrplus}. The reward for refining the element $K_j$ in the moved mesh $\widetilde{\mathcal{T}}^{(n)}$ is:
\begin{equation}
\label{element reward with alpha}
 r^{(n), h}_{K_j} = \bigl({\eta_{K_j, \infty}}-\max_{k \in \text{children}(K_j)}{\eta_{K_k, \infty}}\bigr) - \alpha\bigl(|\text{children}(K_j)|-1\bigr).
\end{equation}
Here, $\text{children}(K_j)$ denotes the set of elements created when subdividing $K_j$ by the chosen refinement procedure, and $|\text{children}(K_j)|$ is their count.
The indicator $\eta_{\cdot, \infty}$ is normalized by the total initial error of the episode to ensure a consistent scale. The cost penalty coefficient $\alpha$ is drawn from a log-uniform distribution during training and supplied as a feature to the element agents, allowing the trained policy to adapt to different cost sensitivities at inference time without retraining.

\paragraph{Global reward components and returns.} To prevent overly excessively local behavior and encourage globally beneficial adaptations, local agent returns are blended in a global signal. For an element agent acting on $K_j$, the final return is the average of its local return $J_{K_j}^{(n), h}$ and global return $J^{(n), h}$, which is the average of all element returns at that step:
\begin{equation}
 \hat J^{(n), h}_{K_j} = \tfrac{1}{2} J_{K_j}^{(n), h} + \tfrac{1}{2} J^{(n), h}.
\label{eq:reward_h}
\end{equation}

For vertex agents, the global signal $g^{(n)}$ is the fractional reduction of the total squared $L_2$ error across the mesh due to the relocation from $\mathcal{T}^{(n)}$ to $\widetilde{\mathcal{T}}^{(n)}$:
$$
 g^{(n)} = \frac{\sum_{K \in \mathcal{T}^{(n)}} {\eta_{K, 2}^2} - \sum_{K \in \widetilde{\mathcal{T}}^{(n)}} {\eta_{K, 2}^2}}{\sum_{K \in \mathcal{T}^{(n)}} {\eta_{K, 2}^2}}.
$$
The final vertex return is then a blend of the local return $ \hat J^{(n), r}_{\boldsymbol{z}_i}$ and $g^{(n)}$, and since they both roughly lie in $[-1, 1]$: 
\begin{equation}
\hat J^{(n), r}_{\boldsymbol{z}_i} = \tfrac{1}{2} J^{(n), r}_{\boldsymbol{z}_i} + \tfrac{1}{2} g^{(n)}.
\label{eq:reward_r}
\end{equation}
This was found to stabilize learning, as averaging the raw rewards directly proved too noisy. 

\paragraph{Training curriculum.} The learning problem is non-stationary: the refinement policy must act on a mesh that is being deformed by a relocation policy while it is being learned. To stabilize this co-adaptation, we employ a two-phase curriculum. In \emph{Phase I}, the refinement actor-critic is frozen, and elements are split randomly. This allows the relocation policy to converge while being exposed to a wide variety of mesh topologies. In \emph{Phase II}, the refinement policy is unfrozen, and both policies are trained jointly using the full Swarm-PPO update. This staged approach empirically removes learning oscillations and accelerates convergence to a robust, coordinated \(hr\)-policy.

\section{Numerical Experiments and Results}
In this section, we evaluate the proposed HypeR across four different 2D PDEs: the Poisson equation on L-shaped domains, the unsteady heat diffusion problem, where we consider the solution at the last time-instance of a preset trajectory, on square domains, Stokes flow on square domains with multiple rhomboid obstacles, and linear elasticity on L-shaped domains (detailed mathematical formulations of the PDE problems are provided in Appendix~A of \citet{freymuth2024asmrplus}). 
Figure~\ref{fig:example_refinements} illustrates representative HypeR refinements for each equation, showcasing the method's ability to adapt the mesh topology to the specific characteristics of different physical phenomena. The columns correspond to different values of the element-penalty characteristic, denoted as $\alpha$ and defined in equation \eqref{element reward with alpha}, provided to the policy at inference. This value, which we vary to generate Pareto fronts, directly modulates the final mesh density (a lower penalty $\alpha$ corresponds to a denser mesh).
\begin{figure}[t]
  \centering
  \includegraphics[width=\textwidth]{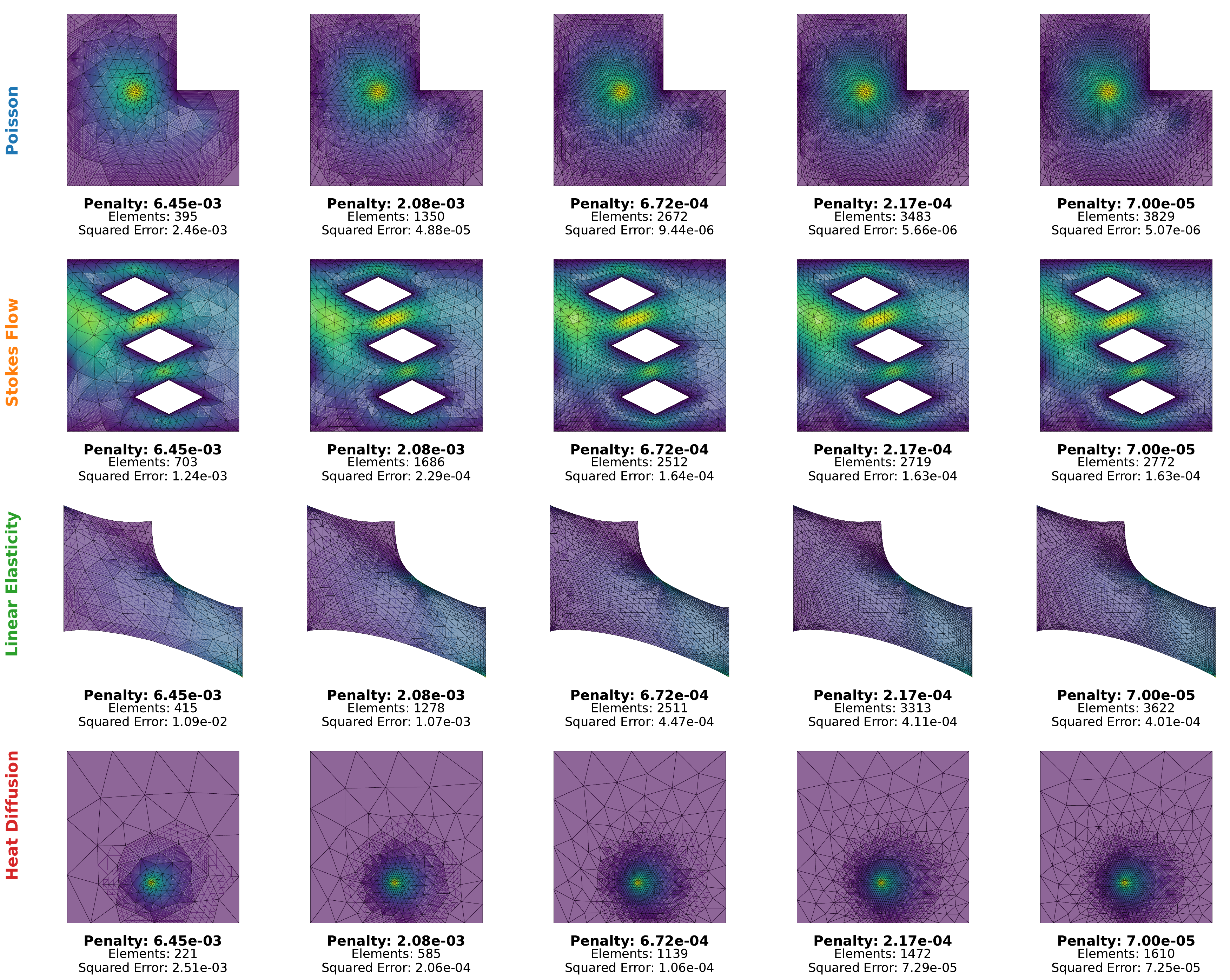}
  \caption{Representative HypeR refinements across four PDE problems (rows) for five penalty settings (columns). At inference time we set an element-penalty feature on hyperedges that controls the cost of adding elements; moving left to right the penalty decreases and the final mesh density increases. This single policy thus can be seen as producing a continuum of meshes by varying a single scalar input.}
  \label{fig:example_refinements}
\end{figure}

\subsection{Experimental Setup}
Our experimental protocol uses, at training time, a fixed dataset of 500 configurations, obtained by employing conforming triangular meshes with linear finite elements and varying the domain geometry and parameters (e.g., load positions and for Poisson/heat diffusion, inlet velocity profiles for Stokes, boundary conditions for the elasticity equation) for each PDE problem, each paired with corresponding initial and reference meshes. Initial meshes are deliberately coarse, containing 20-40 elements for most domains and 80-100 elements for Stokes flow due to the increased complexity of physical phenomenon. Reference meshes $\Omega_{\text{ref}}$, on which the high-fidelity reference solutions are computed, are constructed by uniformly refining the initial mesh 6 times (5 times for Stokes flow due to computational constraints). These high-fidelity solutions are then used for reward computation, as in \eqref{eq:reward_h} and \eqref{eq:reward_r}, during training.

Each refinement strategy operates for 4 adaptation steps across all equations except Stokes flow, which uses 3 steps. Within HypeR's framework, this corresponds to 4 (or 3) combined relocation-refinement operations followed by a final relocation-only step that optimizes the final mesh node locations without adding elements. Training proceeds for 400 PPO iterations with a two-stage curriculum: during the first 150 iterations the relocation policy is trained with frozen refinement (Phase I), followed by 250 iterations of joint training (Phase II) as described in Section~\ref{sec:reward}.

We evaluate model performance using a discrete approximation of the element weighted $L^2$-norm of the error computed on the reference mesh $\Omega_{\text{ref}}$. Let $\mathcal{T}_{\text{ref}}$ be its triangulation with elements $K^*$ of volume $|K^*|$, and let $\mathbf{u}_{\text{ref}}$ denote the FEM solution on $\mathcal{T}_{\text{ref}}$. For a solution $\mathbf{u}_{\mathcal{T}}$ obtained on a mesh $\mathcal{T}$, we evaluate it on the reference mesh via the interpolation operator $\Pi_{\text{ref}}$\footnote{The operator $\Pi_{\text{ref}}$ samples the continuous FEM function $u_{\mathcal{T}}$ at the centroids of the reference mesh elements. Specifically, for a centroid $p_{K^*}$ of a reference element $K^*$, the value $(\Pi_{\text{ref}} u_{\mathcal{T}})(p_{K^*})$ is determined by identifying the coarse element $K \in \mathcal{T}$ such that $p_{K^*} \subset K$ and evaluating the solution using the basis functions associated to $K$ in $p_{K^*}$ points in space.} and form the volume-weighted squared discrepancy:
$$
  \mathcal{E}^2(\mathcal{T}) = \sum_{K^* \in \mathcal{T}_{\text{ref}}} |K^*| \, \big\| \mathbf{u}_{\text{ref}}(p_{K^*}) - (\Pi_{\text{ref}} \mathbf{u}_{\mathcal{T}})(p_{K^*}) \big\|_{\ell^2}^2,
$$
where $p_{K^*}$ is the centroid of the reference element $K^*$. We report the normalized squared error $\mathcal{E}^2_{\text{rel}}(\mathcal{T}) = \mathcal{E}^2(\mathcal{T})/\mathcal{E}^2(\mathcal{T}_{\text{init}})$ to enable comparison across PDEs. Model performance is assessed on a separate evaluation set of 100 PDE configurations. We evaluate each trained model across 20 different element penalty values $\alpha$ at inference time, producing a spectrum of mesh resolutions for comprehensive analysis. Complete hyperparameter specifications are provided in Appendix~\ref{app_sec:hyperparameters}. All experiments are repeated for 3 different random seeds.

\subsection{Baselines and Comparisons}

We compare HypeR against both learning-based and traditional adaptive mesh refinement methods. For learning-based comparisons, we evaluate our approach against ASMR++ \cite{freymuth2024asmrplus}, the current state-of-art deep RL approach for adaptive mesh refinement. ASMR++ employs a swarm-based multi-agent framework for $h$-adaptive refinement but lacks the $r$-adaptive vertex relocation capabilities of HypeR. We use identical experimental conditions for fair comparison: the same refinement steps, training and evaluation datasets, domain, and PDE formulations. We omit comparisons with other deep learning methods such as VDGN \cite{yang2023multiagent}, Single Agent \cite{yang2023reinforcement}, and Sweep \cite{foucart2023deep}, as ASMR++ has demonstrated superior performance against these baselines across similar problem settings.

For traditional heuristic methods, we compare HypeR against the maximum Error Oracle heuristic \cite{dorfler1996convergent}, which iteratively refines elements based on the maximum pointwise error within each element using oracle information from the reference mesh $\Omega_{\text{ref}}$. We also evaluate the proposed approach against ZZ1 and ZZ2 error estimators, which employ the Zienkiewicz–Zhu \cite{zienkiewicz1992superconvergent} superconvergent patch recovery technique with 1 and 2 initial uniform refinement steps, respectively. The initial refinement steps are necessary as the ZZ error estimator produces unreliable results on very coarse meshes, failing to detect solution gradients in regions where elements are too large relative to the solution features.

All baseline methods operate under the same computational constraints and evaluation metrics as HypeR, enabling direct performance comparisons across the error-element count Pareto frontier.

\subsection{Main Results}
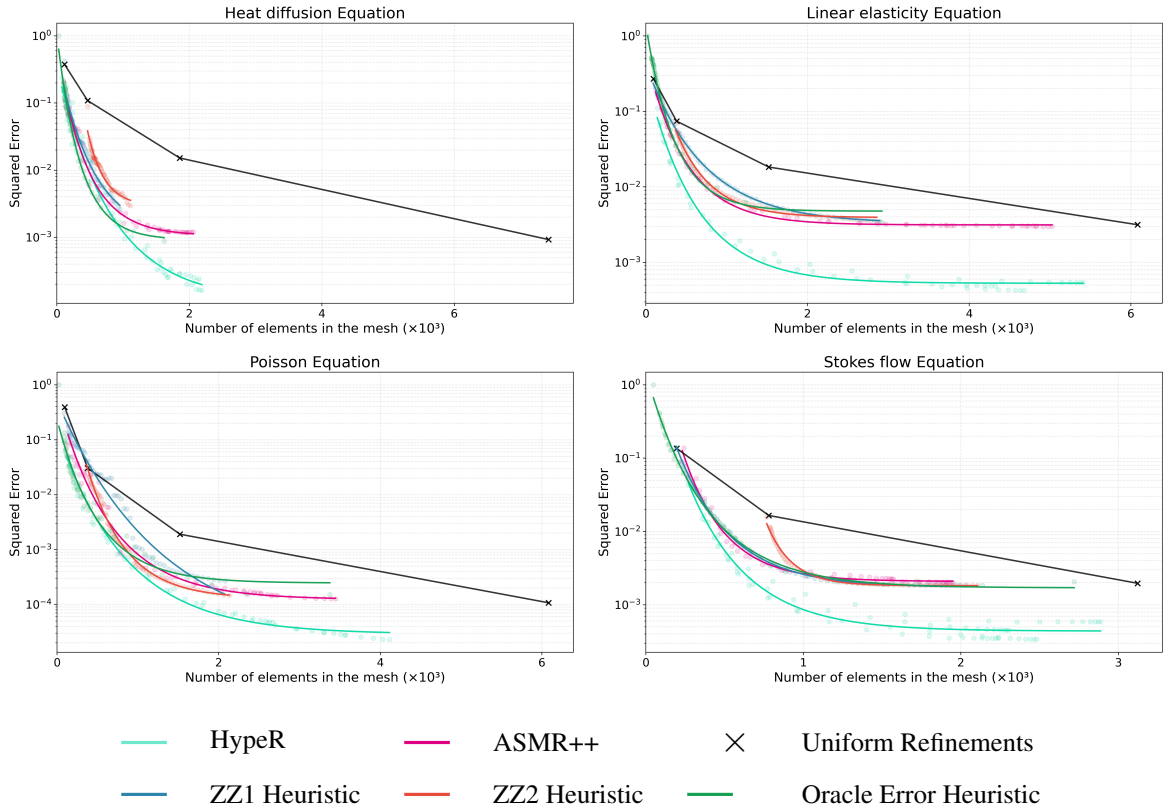
\begin{figure}[t]
  \centering
  \begin{subfigure}[b]{0.45\textwidth}
    \includegraphics[width=\textwidth]{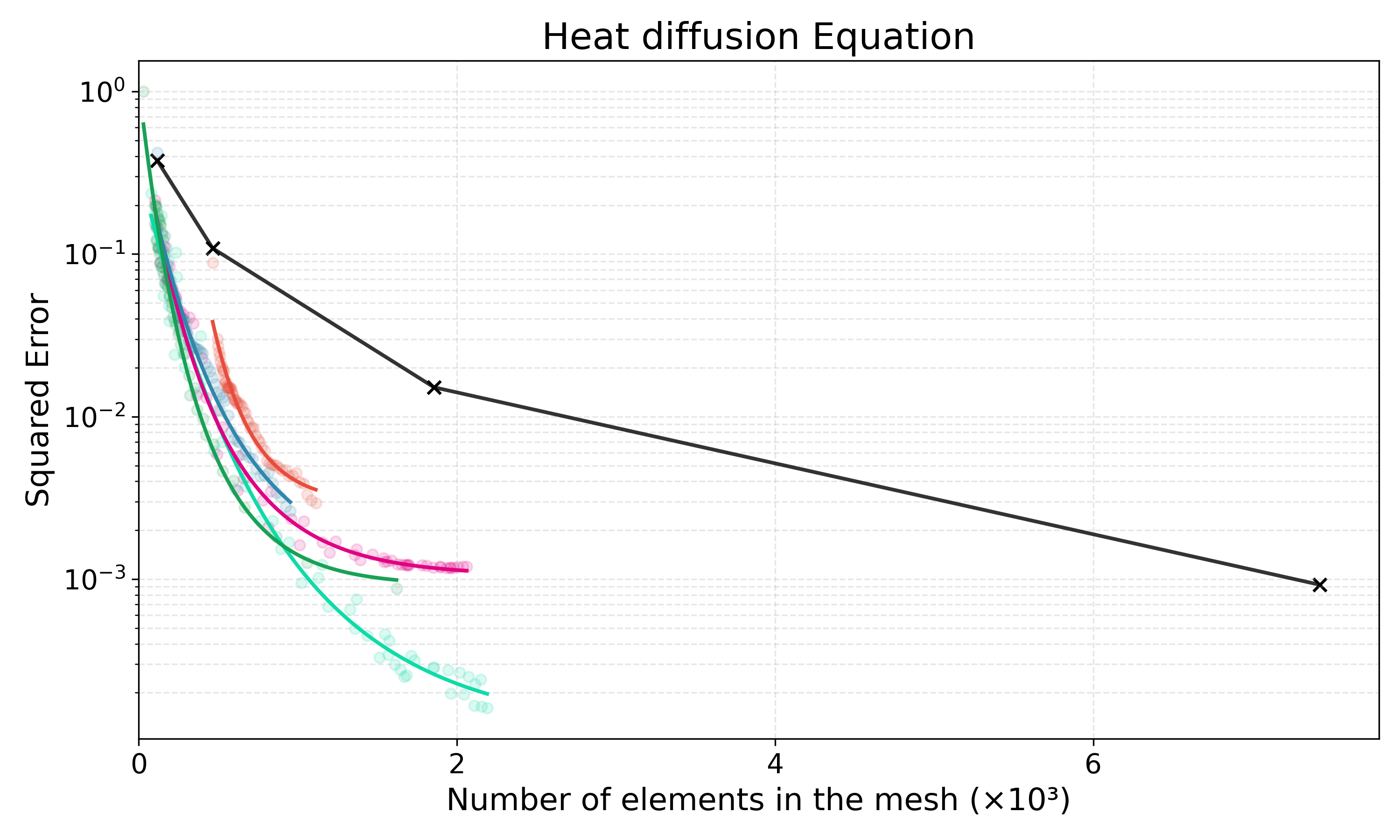}
  \end{subfigure}
  \begin{subfigure}[b]{0.45\textwidth}
    \includegraphics[width=\textwidth]{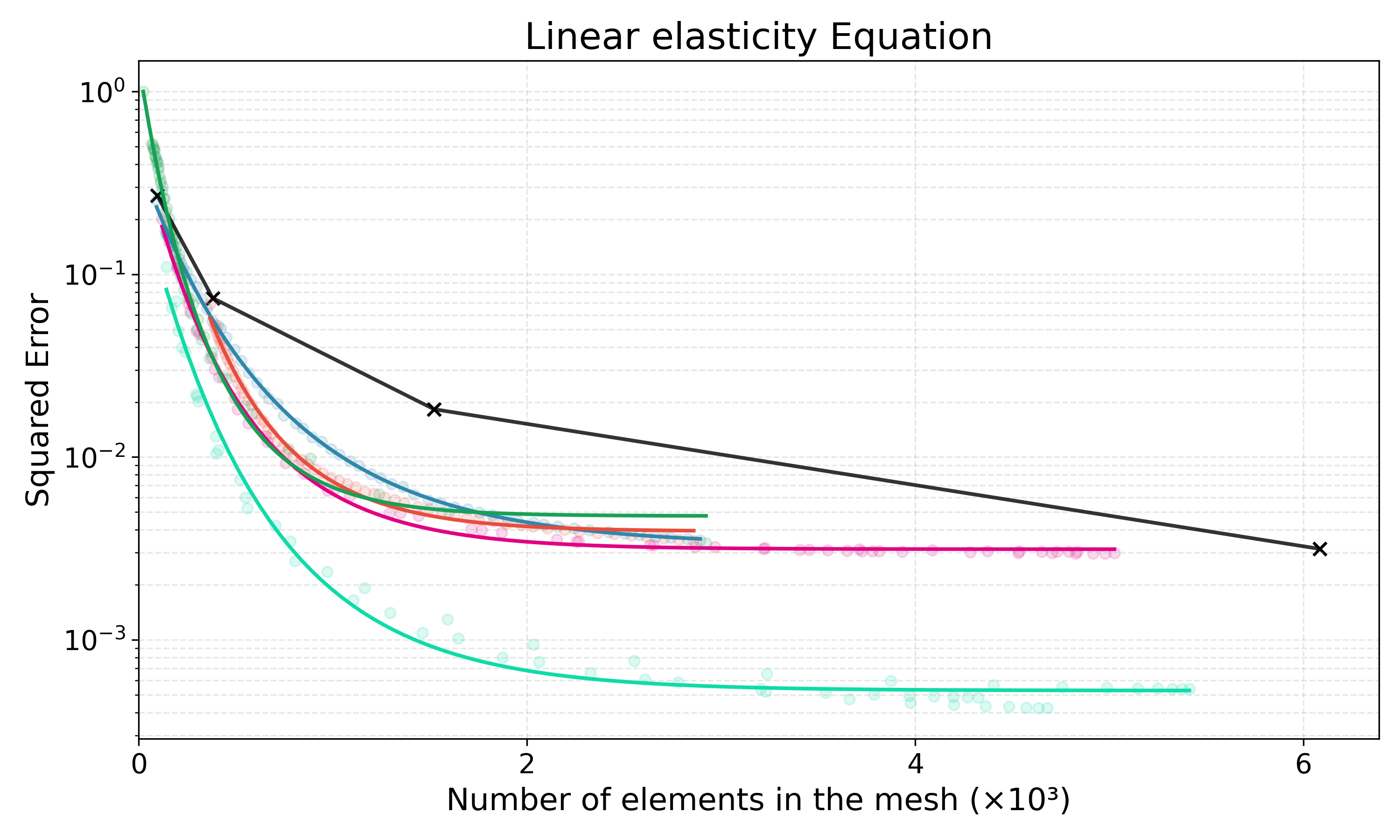}
  \end{subfigure}
  \begin{subfigure}[b]{0.45\textwidth}
    \includegraphics[width=\textwidth]{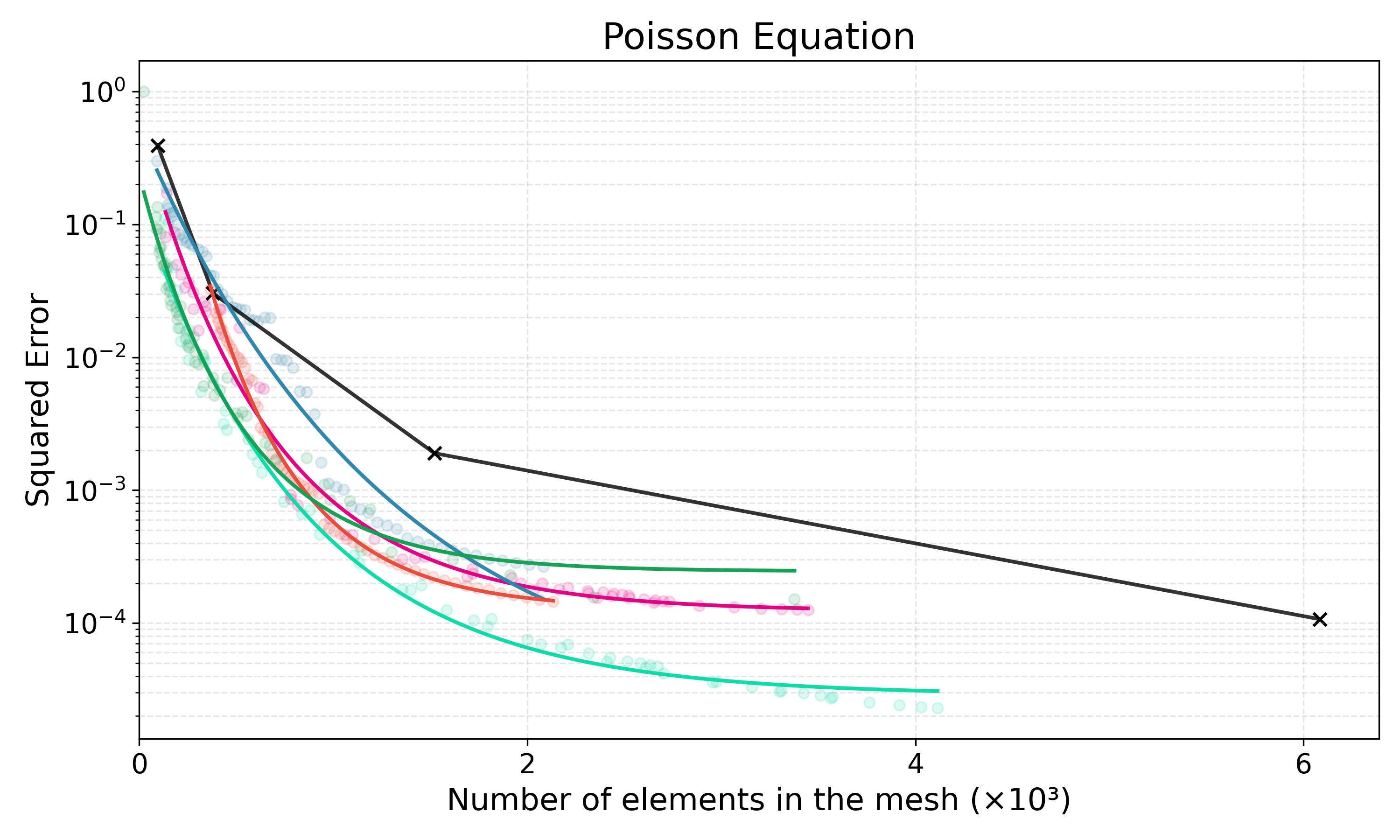}
  \end{subfigure}
  \begin{subfigure}[b]{0.45\textwidth}
    \includegraphics[width=\textwidth]{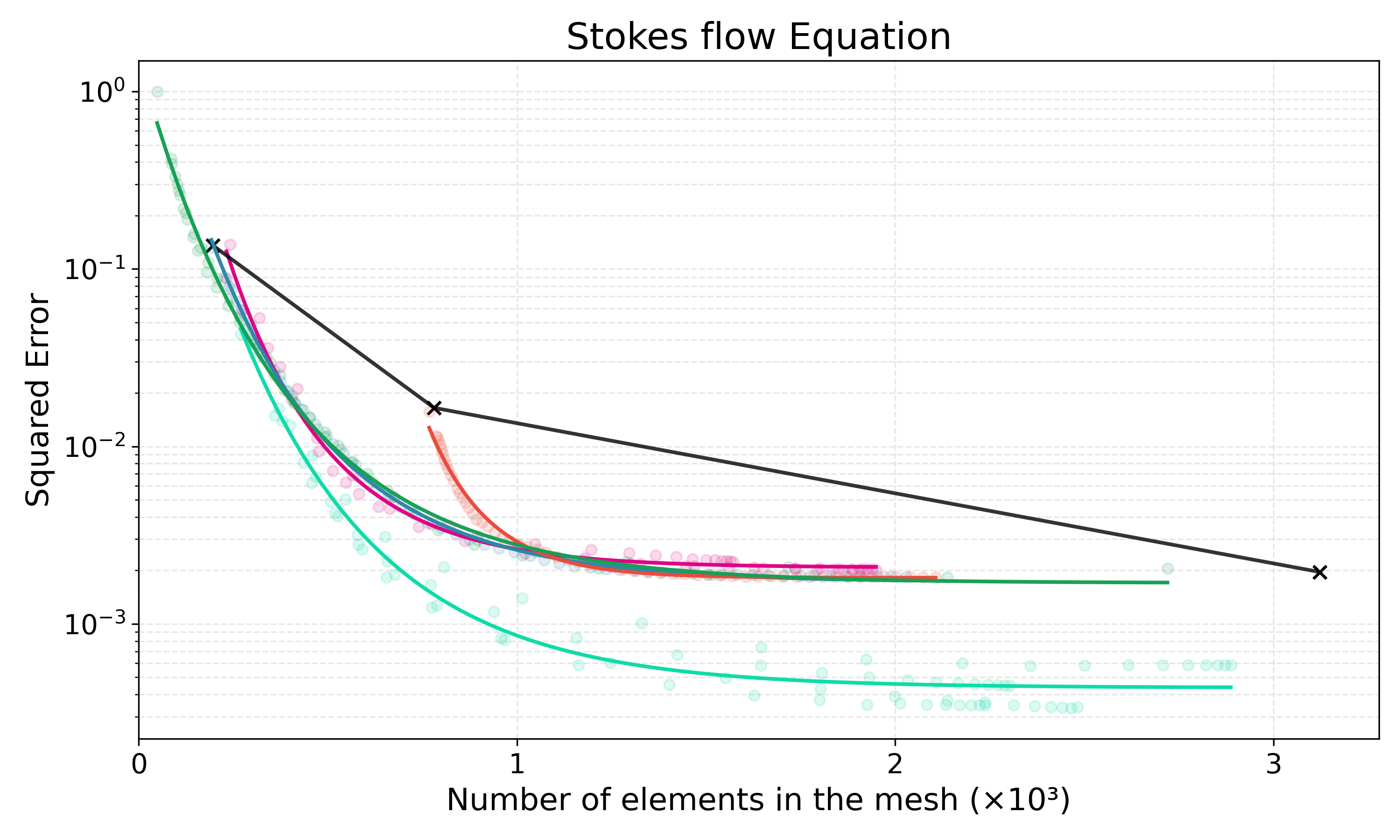}
  \end{subfigure}

\vspace{0.5em}
\begin{tikzpicture}
  \matrix[matrix of nodes,
          nodes={anchor=west, font=\small},
          column sep=12pt, row sep=4pt]{
    \draw[very thick, color={rgb,255:red,12; green,219; blue,167}, opacity=0.6] (0,0) -- (0.6,0); & HypeR &
    \draw[very thick, color={rgb,255:red,224; green,0; blue,131}] (0,0) -- (0.6,0); & ASMR++ &
    \node at (0.3,0) {\Large $\times$}; & Uniform Refinements \\
    \draw[very thick, color={rgb,255:red,46; green,134; blue,171}] (0,0) -- (0.6,0); & ZZ1 Heuristic &
    \draw[very thick, color={rgb,255:red,231; green,76; blue,60}] (0,0) -- (0.6,0); & ZZ2 Heuristic &
    \draw[very thick, color={rgb,255:red,22; green,160; blue,84}] (0,0) -- (0.6,0); & Oracle Error Heuristic \\
  };
\end{tikzpicture}
  \caption{Convergence analysis across four benchmark PDEs showing squared error vs. number of mesh elements. 
  HypeR (water green) consistently achieves lower error than ASMR++ (magenta), ZZ1/ZZ2 (blue/light red), Oracle (green), and uniform refinement (black crosses). Exponential saturation curves fitted to mean performance over 100 test instances.}
  \label{fig:convergence_analysis}
\end{figure}

We evaluate HypeR along two  dimensions: accuracy, or error reduction, and computational efficiency. For accuracy assessment, we employ Pareto curve plots showing the trade-off between element count and normalized squared error relative to the initial mesh. Each point represents the mean performance across our 100-evaluation dataset, with exponential saturation regression curves fitted to capture convergence behavior across the spectrum of mesh densities achieved by varying penalty parameters.

\paragraph{Accuracy analysis.} Figure~\ref{fig:convergence_analysis} demonstrates HypeR's superior performance across all four benchmark problems. On the Poisson equation, HypeR achieves an error approximately 6 times smaller than ASMR++ at comparable element counts, while for linear elasticity, the error reduction reaches a factor of 10. The Stokes flow and heat diffusion problems show consistent improvements, with errors from 3 to 5 times smaller, as HypeR maintains steeper error reduction over the evaluated element ranges with respect to the other methods which quickly attain an error plateau.

A critical insight emerges from the exponential saturation analysis: ASMR++, ZZ1, ZZ2, and Oracle approaches all converge to accuracy limits imposed by uniform mesh refinement constraints. These methods can only subdivide existing elements, creating an inherent ceiling around $10^{-3}$ to $10^{-4}$ normalized error, corresponding to the best accuracy achievable through only element subdivision. HypeR fundamentally overcomes this limitation through its joint relocation-refinement approach, achieving error levels of $10^{-4}$ to $10^{-5}$ while using $2-3\times$ fewer elements than uniform refinement. This joint capability allows HypeR to simultaneously modify the mesh topology (via subdivision) and optimize its geometry (via vertex relocation) to better align with solution features. 

\paragraph{Computational efficiency and robustness}
To provide a direct statistical comparison complementary to the Pareto curves in Figure~\ref{fig:convergence_analysis}, Table~\ref{tab:performance_comparison} aggregates performance over specific intervals of element counts (e.g., $2500\text{--}3500$ elements for Poisson). Since adaptive strategies dictate the final mesh size dynamically, they rarely produce meshes with identical element counts. Therefore, we define a representative target interval for each PDE and compute the mean error with standard deviation and runtime for all evaluation episodes where the final mesh size falls within this range.
This binning approach allows us to compare several methods average performance under similar resolution constraints. Within these intervals, HypeR consistently achieves the lowest mean error, outperforming the strongest baseline (ASMR++) by factors ranging from $2.9\times$ (Poisson) to over $6\times$ (linear elasticity).
Regarding computational costs, while the joint \(hr\)-formulation introduces a modest overhead relative to ASMR++ ($1.5\text{--}2\times$ inference time) due to the additional relocation head and intermediate solver step, HypeR remains orders of magnitude faster than the Oracle heuristic and competitive with ZZ estimators. This confirms that the method delivers a superior trade-off, converting a small investment in inference compute into substantial gains in approximation accuracy.

\begin{table}[htbp]
\centering
\renewcommand{\arraystretch}{1.2}
\setlength{\tabcolsep}{4pt}
\resizebox{\textwidth}{!}{
\begin{tabular}{|l|c|ccccc|ccccc|}
\hline
\multirow{2}{*}{PDE} 
& \multirow{2}{*}{\makecell{Elements\\Range}} 
& \multicolumn{5}{c|}{Mean Time (s)} 
& \multicolumn{5}{c|}{Mean Error} \\ \cline{3-12}
&  
& HypeR & ASMR++ & Oracle & ZZ1 & ZZ2
& HypeR & ASMR++ & Oracle & ZZ1 & ZZ2 \\ \hline

Poisson          
& 2500-3500 
& \makecell{0.209\\{\scriptsize ±0.029}} & \makecell{0.107\\{\scriptsize ±0.017}} & \makecell{2.462\\{\scriptsize ±14.86}} & \makecell{0.979\\{\scriptsize ±5.41}} & \makecell{1.036\\{\scriptsize ±5.61}}
& \makecell{\textbf{4.36e-5}\\{\scriptsize ±4.36e-5}} & \makecell{1.28e-4\\{\scriptsize ±1.16e-4}} & \makecell{1.51e-4\\{\scriptsize ±1.56e-4}} & \makecell{1.07e-4\\{\scriptsize ±8.52e-5}} & \makecell{1.04e-4\\{\scriptsize ±8.71e-5}} \\

Stokes Flow      
& 1000-2000  
& \makecell{0.440\\{\scriptsize ±0.076}} & \makecell{0.274\\{\scriptsize ±0.039}} & \makecell{3.749\\{\scriptsize ±17.93}} & \makecell{1.887\\{\scriptsize ±7.63}} & \makecell{0.779\\{\scriptsize ±3.88}}
& \makecell{\textbf{8.34e-4}\\{\scriptsize ±5.35e-4}} & \makecell{2.32e-3\\{\scriptsize ±9.80e-4}} & \makecell{2.02e-3\\{\scriptsize ±6.43e-4}} & \makecell{1.97e-3\\{\scriptsize ±6.57e-4}} & \makecell{2.01e-3\\{\scriptsize ±6.98e-4}} \\

Linear Elasticity
& 2000-5000 
& \makecell{0.523\\{\scriptsize ±0.120}} & \makecell{0.312\\{\scriptsize ±0.069}} & \makecell{2.130\\{\scriptsize ±10.78}} & \makecell{0.530\\{\scriptsize ±2.97}} & \makecell{0.555\\{\scriptsize ±2.99}}
& \makecell{\textbf{4.95e-4}\\{\scriptsize ±4.09e-4}} & \makecell{3.16e-3\\{\scriptsize ±1.08e-3}} & \makecell{3.53e-3\\{\scriptsize ±8.55e-4}} & \makecell{3.41e-3\\{\scriptsize ±1.03e-3}} & \makecell{3.35e-3\\{\scriptsize ±1.00e-3}} \\

Heat Diffusion   
& 1300-2000  
& \makecell{0.278\\{\scriptsize ±0.039}} & \makecell{0.167\\{\scriptsize ±0.021}} & \makecell{3.746\\{\scriptsize ±12.32}} & \makecell{0.699\\{\scriptsize ±2.74}} & \makecell{0.366\\{\scriptsize ±0.77}}
& \makecell{\textbf{3.61e-4}\\{\scriptsize ±5.62e-4}} & \makecell{1.32e-3\\{\scriptsize ±1.61e-3}} & \makecell{1.06e-3\\{\scriptsize ±9.85e-4}} & \makecell{1.45e-3\\{\scriptsize ±1.91e-3}} & \makecell{1.52e-3\\{\scriptsize ±2.19e-3}} \\ \hline
\end{tabular}
}
\caption{Performance comparison across four PDE benchmarks. HypeR achieves the lowest error (bold) across all problems while maintaining competitive computational times compared to traditional heuristic methods.}
\label{tab:performance_comparison}
\end{table}

\begin{figure}[t]
  \centering
  \includegraphics[width=\textwidth]{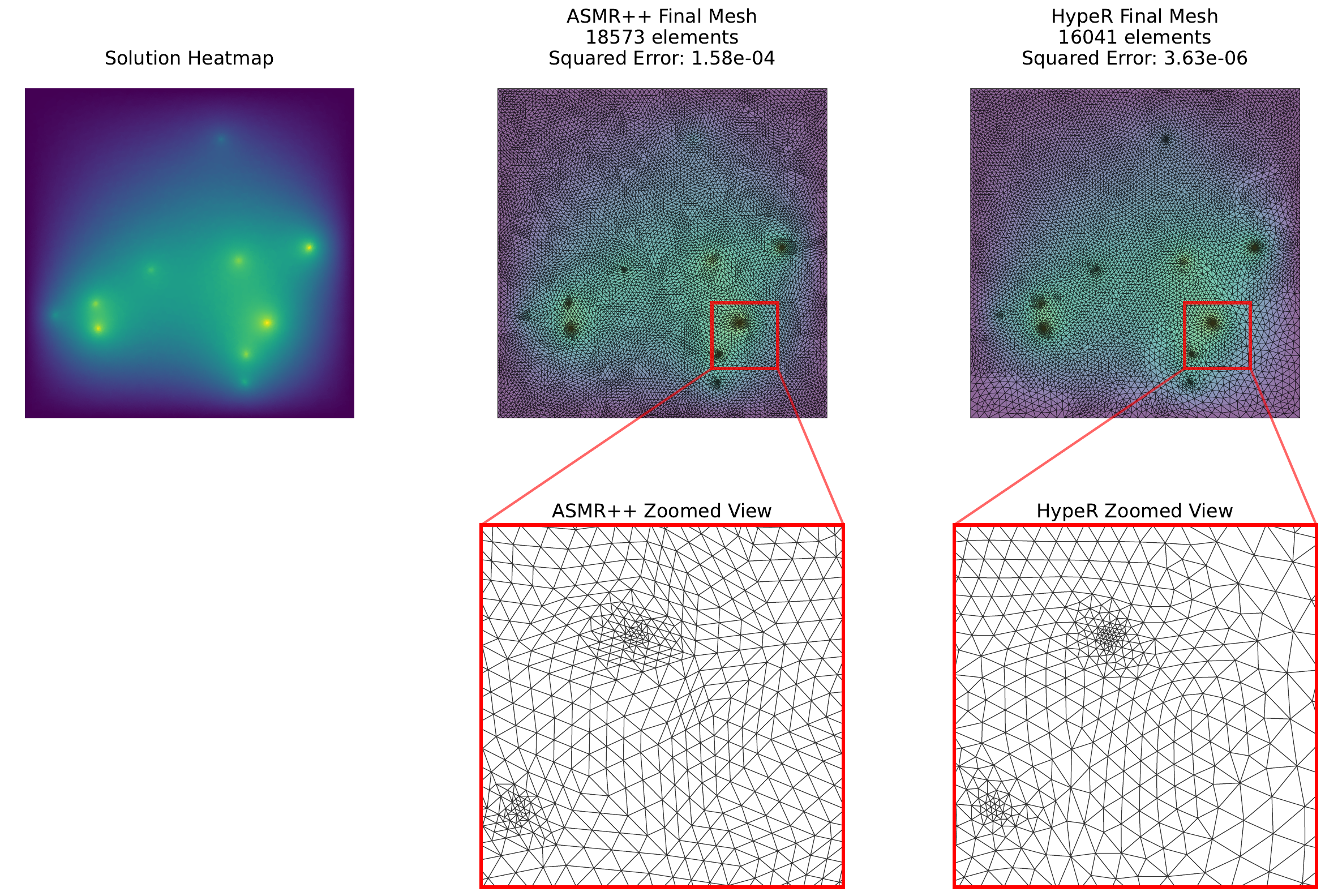}
  \caption{Generalization to much larger domains for Poisson equation (the domain is a 6$\times$6 square versus 1$\times$1 in training). Top: solution field with final mesh overlays for HypeR and ASMR++; bottom: matched zooms. HypeR preserves alignment with solution level sets and smooth mesh gradation, while ASMR++ remains coarser and less smooth in density transitions, leading to higher local error precisely where increased resolution is needed.}
  \label{fig:generalization_big_domain}
\end{figure}

\paragraph{Generalization tests.} Beyond in-distribution testing, we assess zero-shot generalization along two directions. First, we scale the Poisson problem from the unit square domain used during training to a much larger $\Omega =  [6, 6 ]$ (36$\times$ larger area) and apply both HypeR and ASMR++ without any retraining. Figure~\ref{fig:generalization_big_domain} shows that HypeR maintains high mesh quality: elements remain well-shaped with smooth density transitions and align with level curves of the solution, concentrating resolution where gradients are highest. In contrast, ASMR++ produces significantly coarser meshes in critical regions, exhibits less smooth density transitions, and yields visibly higher local error in the zoomed panels where accuracy is most needed. These results indicate that HypeR's joint relocation-refinement policy is robustly transferred to substantially larger domains. 

\begin{figure}[t]
  \centering
  \includegraphics[width=\textwidth]{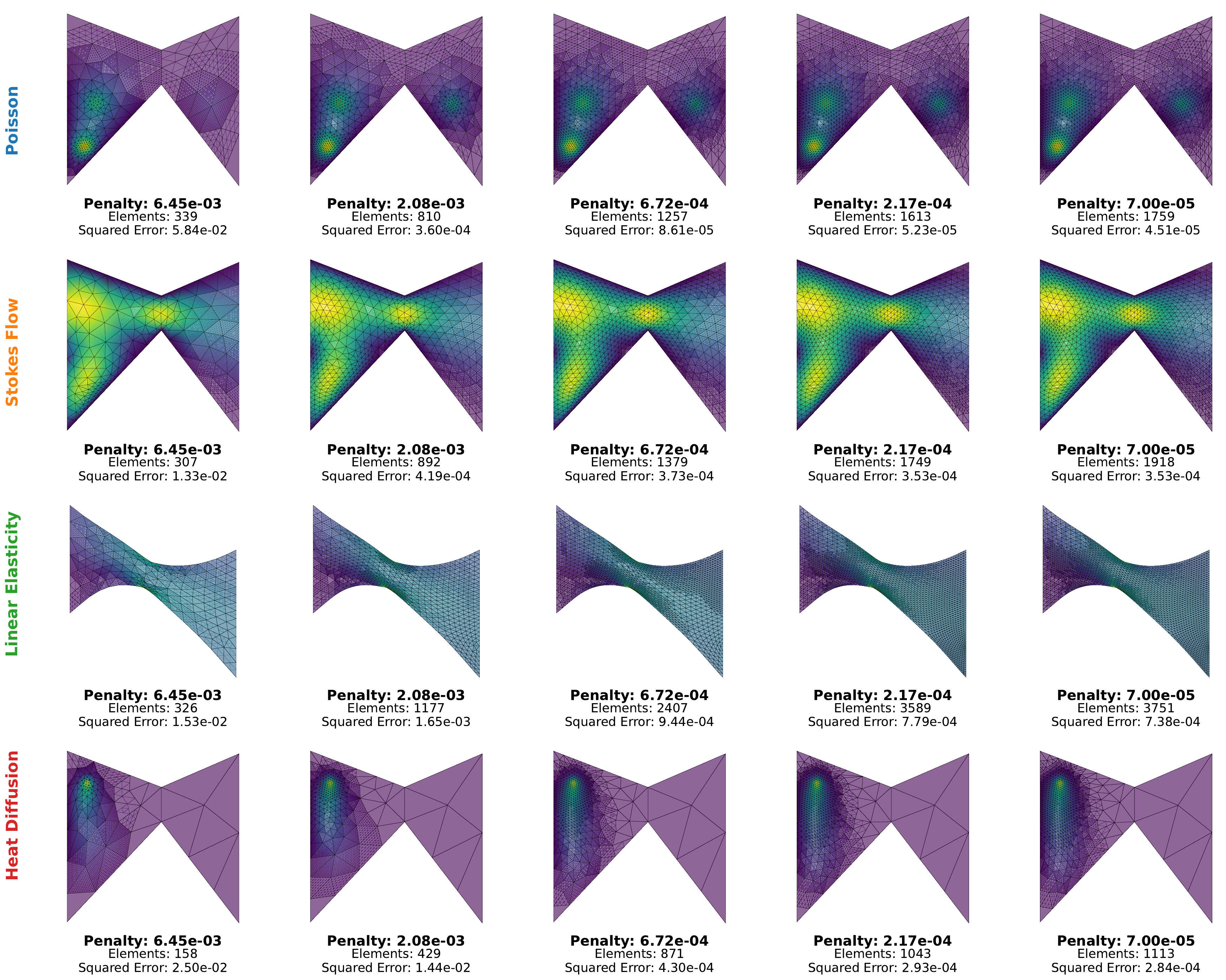}
  \caption{Generalization to different domain shape (unseen during training): a bow-tie geometry with a narrow bottleneck shared across all PDEs. Final meshes and solutions across penalty settings are shown. HypeR retains smooth density transitions, solution-aligned anisotropy, and well-shaped elements without tangling across PDE equations, indicating robust cross-geometry transfer.}
  \label{fig:generalization_bottleneck}
\end{figure}

Second, we evaluate cross-geometry generalization on an unseen "bow-tie" domain with a narrow bottleneck in the center, shared across all equations. Figure~\ref{fig:generalization_bottleneck} (grid across penalties) illustrates that HypeR adapts effectively to this out-of-distribution shape, producing meshes with the same desirable properties observed in-distribution: smooth gradation, alignment to solution anisotropy through relocation, and well-shaped elements without tangling. Performance trends remain consistent across PDEs and penalty settings, demonstrating that the learned policy is not specialized to the training domains.

\begin{figure}[t]
\centering
\begin{minipage}{0.80\textwidth}
  \centering
  \includegraphics[width=\textwidth]{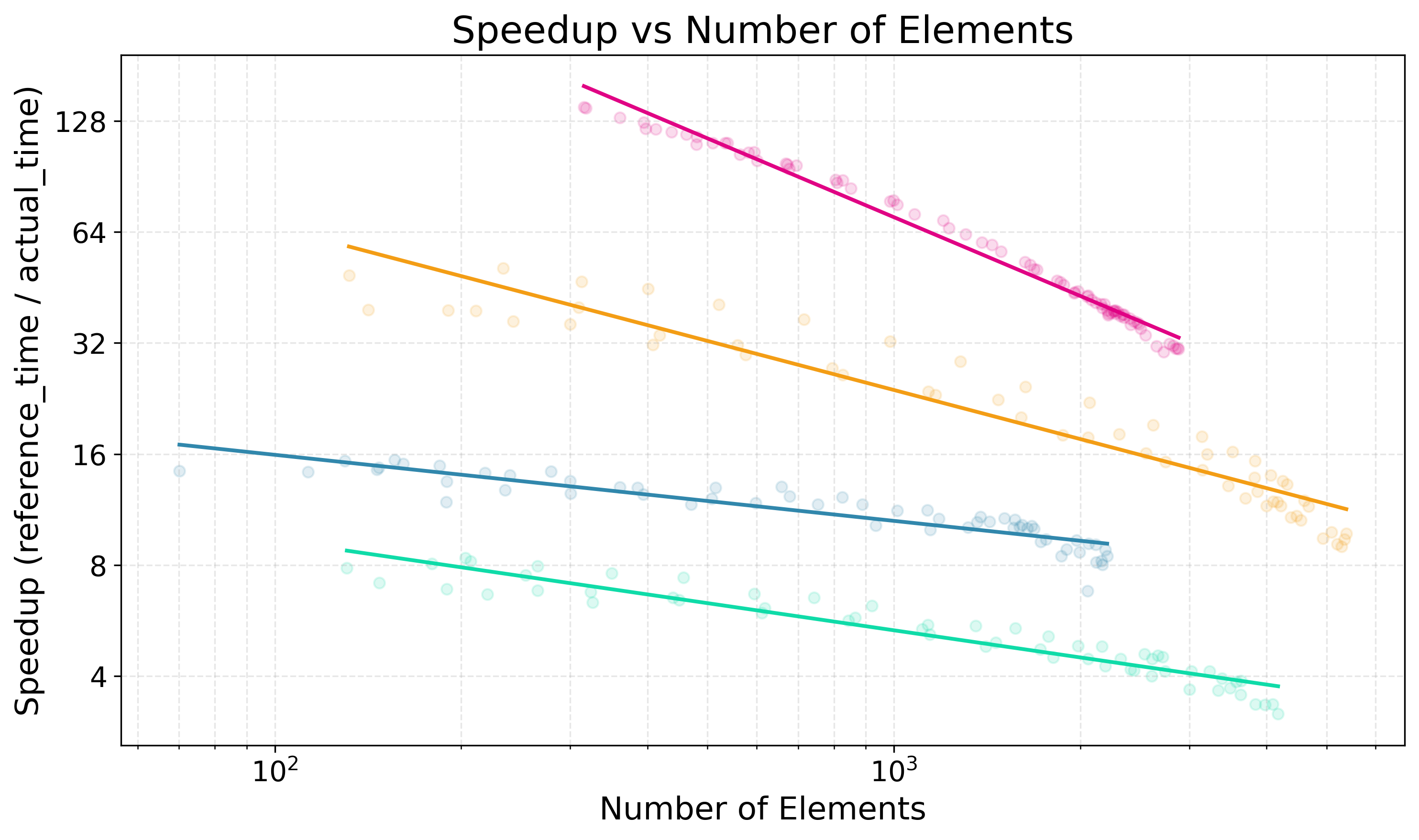}
\end{minipage}%
\hspace{1em}
\begin{minipage}{0.15\textwidth}
  \centering
  \begin{tikzpicture}
    \matrix[matrix of nodes,
            nodes={anchor=west, font=\small},
            row sep=6pt]{
      \draw[very thick, color={rgb,255:red,12; green,219; blue,167}, opacity=0.6] (0,0) -- (0.6,0); & Poisson \\
      \draw[very thick, color={rgb,255:red,224; green,0; blue,131}] (0,0) -- (0.6,0); & Stokes Flow \\
      \draw[very thick, color={rgb,255:red,243; green,156; blue,18}] (0,0) -- (0.6,0); & Linear Elasticity \\
      \draw[very thick, color={rgb,255:red,46; green,134; blue,171}] (0,0) -- (0.6,0); & Heat Diffusion \\
    };
  \end{tikzpicture}
\end{minipage}
\caption{Speedup relative to computing the reference solution. We report the$\text{speedup} = T_{\text{solve}}(\Omega_{\text{ref}}) /  T_{\text{solve}}(\Omega_{\text{HypeR}})$ as a function of the final element count of the HypeR mesh. Each point represents the mean speedup over 100  problem's configurations for a specific mesh density (controlled by the penalty $\alpha$), demonstrating that HypeR is orders of magnitude faster than obtaining the equivalent ground-truth solution via uniform refinement.}
\label{fig:speedup}
\end{figure}

\paragraph{Reference mesh speedup.} The computational benefits become most evident when comparing HypeR against the cost of obtaining the high-fidelity reference solution. We define the speedup as the ratio between the runtime of solving the PDE on the reference mesh $\Omega_{\text{ref}}$, constructed by performing $N+1$ uniform subdivisions of the initial domain, where $N$ is the maximum refinement depth of the HypeR policy, and the total runtime entailed by the HypeR framework (comprising neural network inference, mesh updates, and the final FEM solve). As shown in Figure \ref{fig:speedup}, HypeR achieves speedups ranging from $4\times$ to over $100\times$. This advantage is particularly pronounced for complex physics like Stokes flow and linear elasticity, where the use of FEM makes the brute-force uniform refinement of $\Omega_{\text{ref}}$ prohibitively expensive compared to the targeted allocation of degrees of freedom by HypeR.

\begin{figure}[t]
  \centering
  \includegraphics[width=\textwidth]{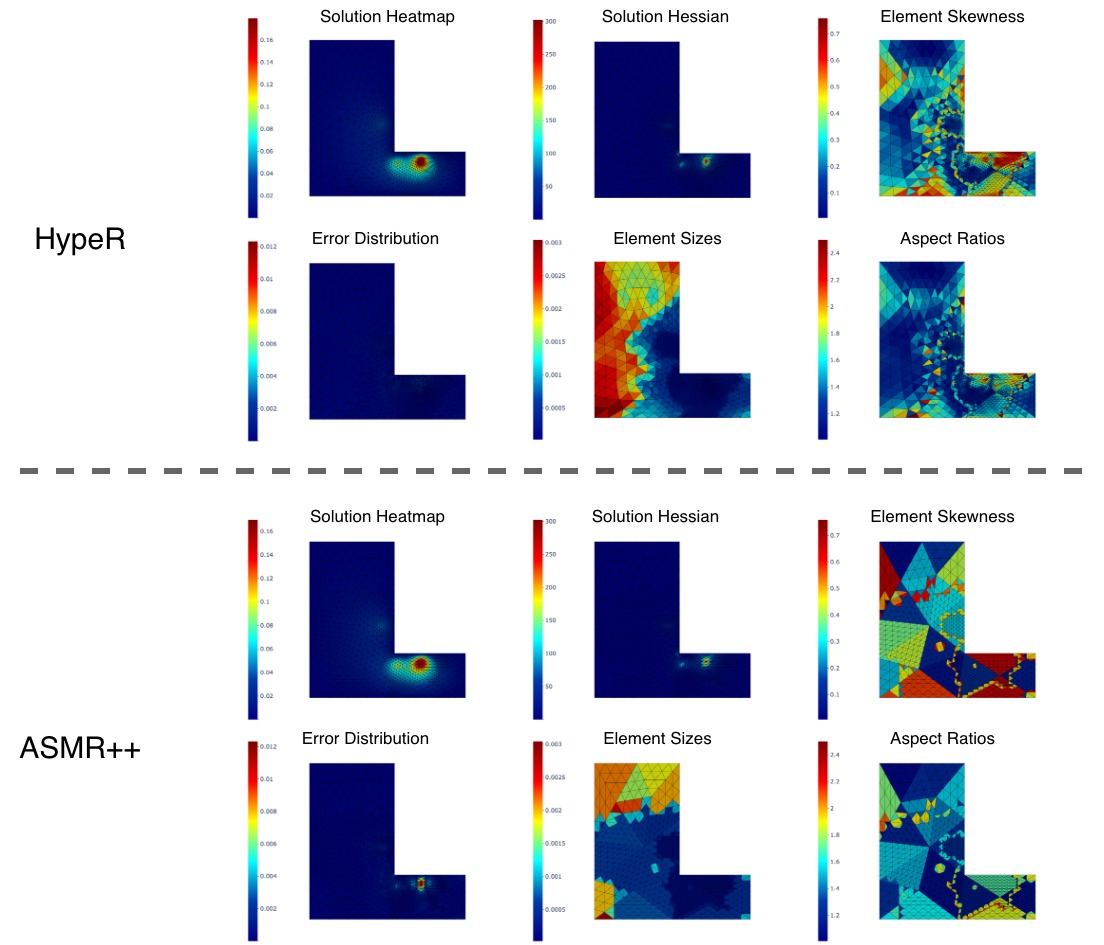}
    \caption{Comparison of mesh quality metrics between ASMR++ (bottom) and HypeR (top) tested on the Poisson equation. HypeR demonstrates superior performance highlighted by smoother transitions in both aspect ratio and skewness across elements, while ASMR++ exhibits abrupt changes and regional concentrations of these properties. HypeR's element size gradually decreases in regions of high solution curvature, in contrast to ASMR++'s abrupt dimensional changes due to hard refinement constraints. Additionally, HypeR's mesh exhibits improved alignment with the solution's contour regions.
    }
  \label{fig:mesh_quality_comparison}
\end{figure}

\paragraph{Qualitative mesh analysis.} Beyond quantitative performance metrics, HypeR produces meshes with demonstrably superior geometric properties compared to refinement-only approaches. Figure~\ref{fig:mesh_quality_comparison} illustrates this through a detailed comparison on the Poisson equation, where HypeR generates meshes with smoother transitions in element size, aspect ratio, and skewness as they approach regions of high solution curvature. While ASMR++ exhibits abrupt dimensional changes due to hard refinement constraints, HypeR's continuous relocation capability enables gradual mesh density variations that better follow the underlying solution structure.
The topology generated by HypeR shows improved alignment with the solution's contour regions, creating element boundaries that naturally follow level curves of the physical field. This geometric coherence translates directly into computational efficiency: well-aligned meshes reduce numerical diffusion, improve condition number of the resulting linear system, and enable more accurate gradient computations with fewer DOFs. The smooth element-size grading also mitigates issues commonly associated with abrupt refinement interfaces - i.e., sharp element-size jumps across neighboring cells - which tend to introduce poorly shaped elements (e.g., extreme aspect ratios, small angles, or high skewness) and numerical artifacts at refinement interfaces.
These qualitative improvements complement the quantitative accuracy gains, demonstrating that HypeR's joint optimization approach produces not merely smaller errors, but fundamentally better-structured computational domains. The resulting meshes exhibit the hallmarks of expert-designed adaptations, that is smooth density transitions, solution-aligned topology, and optimal element quality distributions, achieved automatically through the learned policy. 
These geometric improvements validate our core hypothesis that joint $hr$-optimization fundamentally outperforms sequential approaches, not just in numerical metrics but in producing meshes that better capture the underlying physics.

\begin{figure}[t]
  \centering
  \includegraphics[width=\textwidth]{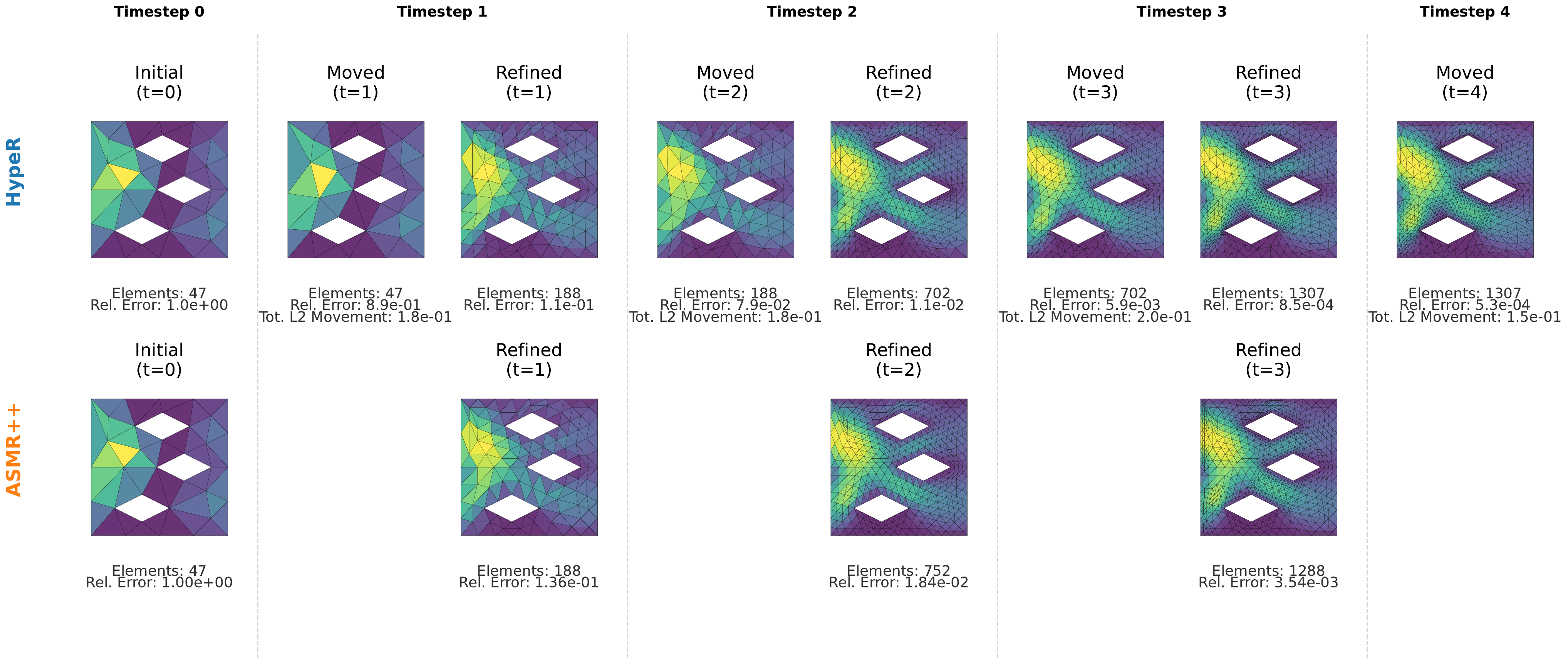}
  \caption{Step-by-step mesh evolution for HypeR (top row) and ASMR++ (bottom row) on a representative Stokes flow problem. 
  Each panel shows the mesh at an adaptation iteration along with corresponding element counts, relative $L^2$ error as defined in the Experimental Setup, and (for HypeR) 
  total vertex displacement. HypeR’s relocation steps produce significant error reduction even before refinement, 
  yielding a final mesh with substantially lower error at comparable resolution. ASMR++ reduces error only through 
  refinement, without geometric optimization.}
  \label{fig:mesh_evolution_stokes}
\end{figure}

\paragraph{Algorithmic evolution.} To further illustrate the mechanisms behind HypeR's quantitative gains, 
Figure~\ref{fig:mesh_evolution_stokes} compares the step-by-step adaptation of HypeR and ASMR++ 
on the Stokes flow problem. For HypeR (top row), relocation 
and refinement are jointly executed at each iteration, followed by a final relocation-only step. This coordination yields substantial error reduction 
at every stage: relocation alone decreases error significantly before refinement even occurs, and the combined 
updates lead to a final mesh with evident lower error at a comparable number of elements. By contrast, 
ASMR++ (bottom row) performs only refinement steps, which steadily reduce error but cannot exploit vertex movement 
to precondition the mesh geometry. The annotated plots beneath each mesh report element counts, relative errors, 
and cumulative vertex displacement, quantifying how relocation accelerates convergence. This visualization 
highlights the qualitative difference between the two algorithms: HypeR actively reshapes the mesh to align with 
solution features, while ASMR++ relies solely on subdivision.

\section{Conclusions and Future Work}

This work introduces HypeR, the first deep RL framework that jointly learns mesh relocation and refinement policies through a unified neural architecture. By computing both continuous vertex displacements and discrete refinement decisions in a single forward pass, HypeR discovers coordinated $hr$-adaptive strategies that fundamentally overcome the limitations of existing approaches. Our framework naturally encompasses pure $h$- and $r$-adaptivity as special cases (one needs to only disable the corresponding policy head), while its joint formulation enables synergies that cannot be achieved by sequential methods. The key insight is that the uniform refinement accuracy ceiling 
can be overcome through learned coordination between topology and geometric modifications.

Our hypergraph neural architecture enables this breakthrough by facilitating bidirectional information flow: element agents communicate refinement demands to vertex agents, which position themselves in advance to ensure well-conditioned subdivisions, while geometric distortion signals from vertex agents trigger targeted topological updates. This coordination, learned through our heterogeneous multi-agent formulation, with distinct discount factors capturing the different temporal impacts of relocation versus refinement, produces meshes achieving a 6--10$\times$ error decrease over state-of-art $h$-adaptive methods, while preserving computational efficiency with respect to both classical heuristics and uniform refinement.

Despite requiring complete FEM solves at each adaptation step and computationally intensive training, due to repeated error computations against reference solutions, HypeR opens promising avenues for accelerating scientific computing workflows. Extension to time-dependent PDEs, while requiring coarsening operations and projection maps for reward assignment on evolving topologies, appears tractable within our framework. Current works explore warm-start techniques and local solve strategies to reduce inference costs, while integration with surrogate models could yield accurate but extremely efficient solution approximations. More broadly, this work demonstrates that many numerical methods addressing $hr$-adaptivity meshing problems through staged procedures may benefit from joint learning approaches that discover coordination strategies beyond human heuristics. As the first method to break the uniform refinement barrier through learned $hr$ coordination, HypeR establishes joint reinforcement-learned adaptivity as a powerful paradigm for automated mesh generation. 


\subsubsection*{Acknowledgments}
SF acknowledges the EPSRC program grant in ``The Mathematics of Deep Learning'', under the project EP/V026259/1. The authors wish to thank Christopher Irwin for his insightful discussions and valuable suggestions which contributed to the conceptual development of this work. Finally, we extend our thanks to Niklas Freymuth and his collaborators for their work on ASMR++ and for making their implementation publicly available; their open-source code provided an invaluable foundation for this project.

\bibliographystyle{cas-model2-names}

\bibliography{references}

@inproceedings{freymuth2023swarm,
  author       = {Niklas Freymuth and
                  Philipp Dahlinger and
                  Tobias W{\"{u}}rth and
                  Simon Reisch and
                  Luise K{\"{a}}rger and
                  Gerhard Neumann},
  editor       = {Alice Oh and
                  Tristan Naumann and
                  Amir Globerson and
                  Kate Saenko and
                  Moritz Hardt and
                  Sergey Levine},
  title        = {Swarm Reinforcement Learning for Adaptive Mesh Refinement},
  booktitle    = {Advances in Neural Information Processing Systems 36: Annual Conference
                  on Neural Information Processing Systems 2023, NeurIPS 2023, New Orleans,
                  LA, USA, December 10 - 16, 2023},
  year         = {2023},
  url          = {http://papers.nips.cc/paper\_files/paper/2023/hash/e85454a113e8b41e017c81875ae68d47-Abstract-Conference.html},
  bibsource    = {dblp computer science bibliography, https://dblp.org}
}

@article{freymuth2024asmrplus,
  author       = {Niklas Freymuth and
                  Philipp Dahlinger and
                  Tobias W{\"{u}}rth and
                  Simon Reisch and
                  Luise K{\"{a}}rger and
                  Gerhard Neumann},
  title        = {Adaptive Swarm Mesh Refinement using Deep Reinforcement Learning with
                  Local Rewards},
  journal      = {CoRR},
  volume       = {abs/2406.08440},
  year         = {2024},
  url          = {https://doi.org/10.48550/arXiv.2406.08440},
  doi          = {10.48550/ARXIV.2406.08440},
  eprinttype   = {arXiv},
  eprint       = {2406.08440},
  bibsource    = {dblp computer science bibliography, https://dblp.org}
}

@inproceedings{yang2023multiagent,
  author       = {Jiachen Yang and
                  Ketan Mittal and
                  Tarik Dzanic and
                  Socratis Petrides and
                  Brendan Keith and
                  Brenden K. Petersen and
                  Daniel M. Faissol and
                  Robert W. Anderson},
  editor       = {Noa Agmon and
                  Bo An and
                  Alessandro Ricci and
                  William Yeoh},
  title        = {Multi-Agent Reinforcement Learning for Adaptive Mesh Refinement},
  booktitle    = {Proceedings of the 2023 International Conference on Autonomous Agents
                  and Multiagent Systems, {AAMAS} 2023, London, United Kingdom, 29 May
                  2023 - 2 June 2023},
  pages        = {14--22},
  publisher    = {{ACM}},
  year         = {2023},
  url          = {https://dl.acm.org/doi/10.5555/3545946.3598614},
  doi          = {10.5555/3545946.3598614},
  bibsource    = {dblp computer science bibliography, https://dblp.org}
}

@inproceedings{yang2023reinforcement,
  author       = {Jiachen Yang and
                  Tarik Dzanic and
                  Brenden K. Petersen and
                  Jun Kudo and
                  Ketan Mittal and
                  Vladimir Z. Tomov and
                  Jean{-}Sylvain Camier and
                  Tuo Zhao and
                  Hongyuan Zha and
                  Tzanio V. Kolev and
                  Robert W. Anderson and
                  Daniel M. Faissol},
  editor       = {Francisco J. R. Ruiz and
                  Jennifer G. Dy and
                  Jan{-}Willem van de Meent},
  title        = {Reinforcement Learning for Adaptive Mesh Refinement},
  booktitle    = {International Conference on Artificial Intelligence and Statistics,
                  25-27 April 2023, Palau de Congressos, Valencia, Spain},
  series       = {Proceedings of Machine Learning Research},
  pages        = {5997--6014},
  publisher    = {{PMLR}},
  year         = {2023},
  url          = {https://proceedings.mlr.press/v206/yang23e.html},
  bibsource    = {dblp computer science bibliography, https://dblp.org}
}

@article{foucart2023deep,
  author       = {Corbin Foucart and
                  Aaron Charous and
                  Pierre F. J. Lermusiaux},
  title        = {Deep reinforcement learning for adaptive mesh refinement},
  journal      = {J. Comput. Phys.},
  volume       = {491},
  pages        = {112381},
  year         = {2023},
  url          = {https://doi.org/10.1016/j.jcp.2023.112381},
  doi          = {10.1016/J.JCP.2023.112381},
  bibsource    = {dblp computer science bibliography, https://dblp.org}
}

@article{dzanic2024dynamo,
  author       = {Tarik Dzanic and
                  Ketan Mittal and
                  DoHyun Kim and
                  J. Yang and
                  Socratis Petrides and
                  Brendan Keith and
                  Robert W. Anderson},
  title        = {DynAMO: Multi-agent reinforcement learning for dynamic anticipatory
                  mesh optimization with applications to hyperbolic conservation laws},
  journal      = {J. Comput. Phys.},
  volume       = {506},
  pages        = {112924},
  year         = {2024},
  url          = {https://doi.org/10.1016/j.jcp.2024.112924},
  doi          = {10.1016/J.JCP.2024.112924},
  bibsource    = {dblp computer science bibliography, https://dblp.org}
}

@article{kim2023gmrnet,
  author       = {Minseong Kim and
                  Jaeseung Lee and
                  Jibum Kim},
  title        = {GMR-Net: GCN-based mesh refinement framework for elliptic {PDE} problems},
  journal      = {Eng. Comput.},
  volume       = {39},
  number       = {5},
  pages        = {3721--3737},
  year         = {2023},
  url          = {https://doi.org/10.1007/s00366-023-01811-0},
  doi          = {10.1007/S00366-023-01811-0},
  bibsource    = {dblp computer science bibliography, https://dblp.org}
}

@article{carstensen2004adaptive,
  title={An adaptive mesh-refining algorithm allowing for an H 1 stable L 2 projection onto Courant finite element spaces},
  author={Carstensen, Carsten},
  journal={Constructive Approximation},
  volume={20},
  number={4},
  pages={549--564},
  year={2004},
  publisher={Springer}
}

@article{yu2024flow2mesh,
  title={Flow2Mesh: A flow-guided data-driven mesh adaptation framework},
  author={Yu, Jian and Lyu, Hongqiang and Xu, Ran and Ouyang, Wenxuan and Liu, Xuejun},
  journal={Physics of Fluids},
  volume={36},
  number={3},
  year={2024},
  publisher={AIP Publishing}
}

@article{bohn2021recurrent,
  author       = {Jan Bohn and
                  Michael Feischl},
  title        = {Recurrent neural networks as optimal mesh refinement strategies},
  journal      = {Comput. Math. Appl.},
  volume       = {97},
  pages        = {61--76},
  year         = {2021},
  url          = {https://doi.org/10.1016/j.camwa.2021.05.018},
  doi          = {10.1016/J.CAMWA.2021.05.018},
  bibsource    = {dblp computer science bibliography, https://dblp.org}
}

@inproceedings{rowbottom2025gadaptivity,
  author       = {James Rowbottom and
                  Georg Maierhofer and
                  Teo Deveney and
                  Eike Hermann M{\"{u}}ller and
                  Alberto Paganini and
                  Katharina Schratz and
                  Pietro Lio and
                  Carola{-}Bibiane Sch{\"{o}}nlieb and
                  Chris J. Budd},
  editor       = {Aarti Singh and
                  Maryam Fazel and
                  Daniel Hsu and
                  Simon Lacoste{-}Julien and
                  Felix Berkenkamp and
                  Tegan Maharaj and
                  Kiri Wagstaff and
                  Jerry Zhu},
  title        = {G-Adaptivity: optimised graph-based mesh relocation for finite element
                  methods},
  booktitle    = {Forty-second International Conference on Machine Learning, {ICML}
                  2025, Vancouver, BC, Canada, July 13-19, 2025},
  series       = {Proceedings of Machine Learning Research},
  publisher    = {{PMLR} / OpenReview.net},
  year         = {2025},
  url          = {https://proceedings.mlr.press/v267/rowbottom25a.html},
  bibsource    = {dblp computer science bibliography, https://dblp.org}
}

@inproceedings{song2022m2n,
  author       = {Wenbin Song and
                  Mingrui Zhang and
                  Joseph G. Wallwork and
                  Junpeng Gao and
                  Zheng Tian and
                  Fanglei Sun and
                  Matthew D. Piggott and
                  Junqing Chen and
                  Zuoqiang Shi and
                  Xiang Chen and
                  Jun Wang},
  editor       = {Sanmi Koyejo and
                  S. Mohamed and
                  A. Agarwal and
                  Danielle Belgrave and
                  K. Cho and
                  A. Oh},
  title        = {{M2N:} Mesh Movement Networks for {PDE} Solvers},
  booktitle    = {Advances in Neural Information Processing Systems 35: Annual Conference
                  on Neural Information Processing Systems 2022, NeurIPS 2022, New Orleans,
                  LA, USA, November 28 - December 9, 2022},
  year         = {2022},
  url          = {http://papers.nips.cc/paper\_files/paper/2022/hash/2f88d8061f12abae9d14d376fd69c933-Abstract-Conference.html},
  bibsource    = {dblp computer science bibliography, https://dblp.org}
}

@inproceedings{zhang2024towards,
  author       = {Mingrui Zhang and
                  Chunyang Wang and
                  Stephan C. Kramer and
                  Joseph G. Wallwork and
                  Siyi Li and
                  Jiancheng Liu and
                  Xiang Chen and
                  Matthew D. Piggott},
  editor       = {Amir Globersons and
                  Lester Mackey and
                  Danielle Belgrave and
                  Angela Fan and
                  Ulrich Paquet and
                  Jakub M. Tomczak and
                  Cheng Zhang},
  title        = {Towards Universal Mesh Movement Networks},
  booktitle    = {Advances in Neural Information Processing Systems 38: Annual Conference
                  on Neural Information Processing Systems 2024, NeurIPS 2024, Vancouver,
                  BC, Canada, December 10 - 15, 2024},
  year         = {2024},
  url          = {http://papers.nips.cc/paper\_files/paper/2024/hash/1b0da24d136f46bfaee78e8da907127e-Abstract-Conference.html},
  bibsource    = {dblp computer science bibliography, https://dblp.org}
}

@article{wang2025ugm2n,
  author       = {Zhichao Wang and
                  Xinhai Chen and
                  Qinglin Wang and
                  Xiang Gao and
                  Qingyang Zhang and
                  Menghan Jia and
                  Xiang Zhang and
                  Jie Liu},
  title        = {{UGM2N:} An Unsupervised and Generalizable Mesh Movement Network via
                  M-Uniform Loss},
  journal      = {CoRR},
  volume       = {abs/2508.08615},
  year         = {2025},
  url          = {https://doi.org/10.48550/arXiv.2508.08615},
  doi          = {10.48550/ARXIV.2508.08615},
  eprinttype   = {arXiv},
  eprint       = {2508.08615},
  bibsource    = {dblp computer science bibliography, https://dblp.org}
}

@inproceedings{hu2024better,
  author       = {Peiyan Hu and
                  Yue Wang and
                  Zhi{-}Ming Ma},
  title        = {Better Neural {PDE} Solvers Through Data-Free Mesh Movers},
  booktitle    = {The Twelfth International Conference on Learning Representations,
                  {ICLR} 2024, Vienna, Austria, May 7-11, 2024},
  publisher    = {OpenReview.net},
  year         = {2024},
  url          = {https://openreview.net/forum?id=hj9ZuNimRl},
  bibsource    = {dblp computer science bibliography, https://dblp.org}
}

@article{yu2025para2mesh,
  title={Para2Mesh: A dual diffusion framework for moving mesh adaptation},
  author={Jian, YU and Hongqiang, LYU and Ran, XU and Xuejun, LIU and others},
  journal={Chinese Journal of Aeronautics},
  volume={38},
  number={7},
  pages={103441},
  year={2025},
  publisher={Elsevier}
}

@article{freymuth2025amber,
  author       = {Niklas Freymuth and
                  Tobias W{\"{u}}rth and
                  Nicolas Schreiber and
                  Bal{\'{a}}zs Gyenes and
                  Andreas Boltres and
                  Johannes Mitsch and
                  Aleksandar Taranovic and
                  Tai Hoang and
                  Philipp Dahlinger and
                  Philipp Becker and
                  Luise K{\"{a}}rger and
                  Gerhard Neumann},
  title        = {{AMBER:} Adaptive Mesh Generation by Iterative Mesh Resolution Prediction},
  journal      = {CoRR},
  volume       = {abs/2505.23663},
  year         = {2025},
  url          = {https://doi.org/10.48550/arXiv.2505.23663},
  doi          = {10.48550/ARXIV.2505.23663},
  eprinttype   = {arXiv},
  eprint       = {2505.23663},
  bibsource    = {dblp computer science bibliography, https://dblp.org}
}

@inproceedings{khan2024graphmesh,
  author       = {Ainulla Khan and
                  Moyuru Yamada and
                  Abhishek Chikane and
                  Manohar Kaul},
  editor       = {Leonardo Franco and
                  Cl{\'{e}}lia de Mulatier and
                  Maciej Paszynski and
                  Valeria V. Krzhizhanovskaya and
                  Jack J. Dongarra and
                  Peter M. A. Sloot},
  title        = {GraphMesh: Geometrically Generalized Mesh Refinement Using GNNs},
  booktitle    = {Computational Science - {ICCS} 2024 - 24th International Conference,
                  Malaga, Spain, July 2-4, 2024, Proceedings, Part {V}},
  series       = {Lecture Notes in Computer Science},
  pages        = {120--134},
  publisher    = {Springer},
  year         = {2024},
  url          = {https://doi.org/10.1007/978-3-031-63775-9\_9},
  doi          = {10.1007/978-3-031-63775-9\_9},
  bibsource    = {dblp computer science bibliography, https://dblp.org}
}

@inproceedings{zhang2020meshingnet,
  author       = {Zheyan Zhang and
                  Yongxing Wang and
                  Peter K. Jimack and
                  He Wang},
  editor       = {Valeria V. Krzhizhanovskaya and
                  G{\'{a}}bor Z{\'{a}}vodszky and
                  Michael Harold Lees and
                  Jack J. Dongarra and
                  Peter M. A. Sloot and
                  S{\'{e}}rgio Brissos and
                  Jo{\~{a}}o Teixeira},
  title        = {MeshingNet: {A} New Mesh Generation Method Based on Deep Learning},
  booktitle    = {Computational Science - {ICCS} 2020 - 20th International Conference,
                  Amsterdam, The Netherlands, June 3-5, 2020, Proceedings, Part {III}},
  series       = {Lecture Notes in Computer Science},
  pages        = {186--198},
  publisher    = {Springer},
  year         = {2020},
  url          = {https://doi.org/10.1007/978-3-030-50420-5\_14},
  doi          = {10.1007/978-3-030-50420-5\_14},
  bibsource    = {dblp computer science bibliography, https://dblp.org}
}

@article{zhang2021meshingnet3d,
  author       = {Zheyan Zhang and
                  Peter K. Jimack and
                  He Wang},
  title        = {MeshingNet3D: Efficient generation of adapted tetrahedral meshes for
                  computational mechanics},
  journal      = {Adv. Eng. Softw.},
  volume       = {157-158},
  pages        = {103021},
  year         = {2021},
  url          = {https://doi.org/10.1016/j.advengsoft.2021.103021},
  doi          = {10.1016/J.ADVENGSOFT.2021.103021},
  bibsource    = {dblp computer science bibliography, https://dblp.org}
}

@article{huang2021machine,
  author       = {Keefe Huang and
                  Moritz Kr{\"{u}}gener and
                  Alistair Brown and
                  Friedrich Menhorn and
                  Hans{-}Joachim Bungartz and
                  Dirk Hartmann},
  title        = {Machine Learning-Based Optimal Mesh Generation in Computational Fluid
                  Dynamics},
  journal      = {CoRR},
  volume       = {abs/2102.12923},
  year         = {2021},
  url          = {https://arxiv.org/abs/2102.12923},
  eprinttype   = {arXiv},
  eprint       = {2102.12923},
  bibsource    = {dblp computer science bibliography, https://dblp.org}
}

@article{zienkiewicz1992superconvergent,
  title={The superconvergent patch recovery and a posteriori error estimates. Part 1: The recovery technique},
  author={Zienkiewicz, Olgierd Cecil and Zhu, Jian Zhong},
  journal={International Journal for Numerical Methods in Engineering},
  volume={33},
  number={7},
  pages={1331--1364},
  year={1992},
  publisher={Wiley Online Library}
}

@article{dorfler1996convergent,
  title={A convergent adaptive algorithm for Poisson’s equation},
  author={D{\"o}rfler, Willy},
  journal={SIAM Journal on Numerical Analysis},
  volume={33},
  number={3},
  pages={1106--1124},
  year={1996},
  publisher={SIAM}
}

@article{binev2004adaptive,
  author       = {Peter Binev and
                  Wolfgang Dahmen and
                  Ronald A. DeVore},
  title        = {Adaptive Finite Element Methods with convergence rates},
  journal      = {Numerische Mathematik},
  volume       = {97},
  number       = {2},
  pages        = {219--268},
  year         = {2004},
  url          = {https://doi.org/10.1007/s00211-003-0492-7},
  doi          = {10.1007/S00211-003-0492-7},
  bibsource    = {dblp computer science bibliography, https://dblp.org}
}

@article{stevenson2008optimality,
  author       = {Rob P. Stevenson},
  title        = {The completion of locally refined simplicial partitions created by
                  bisection},
  journal      = {Math. Comput.},
  volume       = {77},
  number       = {261},
  pages        = {227--241},
  year         = {2008},
  url          = {https://doi.org/10.1090/S0025-5718-07-01959-X},
  doi          = {10.1090/S0025-5718-07-01959-X},
  bibsource    = {dblp computer science bibliography, https://dblp.org}
}

@article{yano2012optimization,
  title={An optimization-based framework for anisotropic simplex mesh adaptation},
  author={Yano, Masayuki and Darmofal, David L},
  journal={Journal of Computational Physics},
  volume={231},
  number={22},
  pages={7626--7649},
  year={2012},
  publisher={Elsevier}
}

@article{mcrae2018optimal,
  author       = {Andrew T. T. McRae and
                  Colin J. Cotter and
                  Chris J. Budd},
  title        = {Optimal-Transport-Based Mesh Adaptivity on the Plane and Sphere Using
                  Finite Elements},
  journal      = {{SIAM} J. Sci. Comput.},
  volume       = {40},
  number       = {2},
  year         = {2018},
  url          = {https://doi.org/10.1137/16M1109515},
  doi          = {10.1137/16M1109515},
  bibsource    = {dblp computer science bibliography, https://dblp.org}
}

@article{nagarajan2018review,
  title={Conforming to interface structured adaptive mesh refinement: 3D algorithm and implementation},
  author={Nagarajan, Anand and Soghrati, Soheil},
  journal={Computational Mechanics},
  volume={62},
  number={5},
  pages={1213--1238},
  year={2018},
  publisher={Springer}
}

@book{evans2010partial,
  title={Partial differential equations},
  author={Evans, Lawrence C},
  volume={19},
  year={2022},
  publisher={American mathematical society}
}

@book{leveque2007finite,
  title={Finite difference methods for ordinary and partial differential equations: steady-state and time-dependent problems},
  author={LeVeque, Randall J},
  year={2007},
  publisher={SIAM}
}

@book{brenner2008mathematical,
  title={The mathematical theory of finite element methods},
  author={Brenner, Susanne C and Scott, L Ridgway},
  year={2008},
  publisher={Springer}
}

@book{ern2004theory,
  title={Theory and practice of finite elements},
  author={Ern, Alexandre and Guermond, Jean-Luc},
  volume={159},
  year={2004},
  publisher={Springer}
}

@article{babuska1994p,
  title={The p and h-p versions of the finite element method, basic principles and properties},
  author={Babu{\v{s}}ka, Ivo and Suri, Manil},
  journal={SIAM review},
  volume={36},
  number={4},
  pages={578--632},
  year={1994},
  publisher={SIAM}
}

@book{saad2003iterative,
  title={Iterative methods for sparse linear systems},
  author={Saad, Yousef},
  year={2003},
  publisher={SIAM}
}

@article{ainsworth2000posteriori,
  title={A posteriori error estimation in finite element analysis},
  author={Ainsworth, Mark and Oden, J Tinsley},
  journal={Computer methods in applied mechanics and engineering},
  volume={142},
  number={1-2},
  pages={1--88},
  year={1997},
  publisher={Elsevier}
}

@book{huang2011adaptive,
  title={Adaptive moving mesh methods},
  author={Huang, Weizhang and Russell, Robert D},
  volume={174},
  year={2010},
  publisher={Springer Science \& Business Media}
}

@article{alauzet2016decade,
  author       = {Fr{\'{e}}d{\'{e}}ric Alauzet and
                  Adrien Loseille},
  title        = {A decade of progress on anisotropic mesh adaptation for computational
                  fluid dynamics},
  journal      = {Comput. Aided Des.},
  volume       = {72},
  pages        = {13--39},
  year         = {2016},
  url          = {https://doi.org/10.1016/j.cad.2015.09.005},
  doi          = {10.1016/J.CAD.2015.09.005},
  bibsource    = {dblp computer science bibliography, https://dblp.org}
}

@article{budd2009adaptivity,
  title={Adaptivity with moving grids},
  author={Budd, Chris J and Huang, Weizhang and Russell, Robert D},
  journal={Acta Numerica},
  volume={18},
  pages={111--241},
  year={2009},
  publisher={Cambridge University Press}
}

@article{delzanno2008optimal,
  author       = {Gian Luca Delzanno and
                  Luis Chac{\'{o}}n and
                  John M. Finn and
                  Yeo{-}Jin Chung and
                  Giovanni Lapenta},
  title        = {An optimal robust equidistribution method for two-dimensional grid
                  adaptation based on Monge-Kantorovich optimization},
  journal      = {J. Comput. Phys.},
  volume       = {227},
  number       = {23},
  pages        = {9841--9864},
  year         = {2008},
  url          = {https://doi.org/10.1016/j.jcp.2008.07.020},
  doi          = {10.1016/J.JCP.2008.07.020},
  bibsource    = {dblp computer science bibliography, https://dblp.org}
}

@article{bank1983some,
  title={Some refinement algorithms and data structures for regular local mesh refinement},
  author={Bank, Randolph E and Sherman, Andrew H and Weiser, Alan},
  journal={Scientific Computing, Applications of Mathematics and Computing to the Physical Sciences},
  volume={1},
  pages={3--17},
  year={1983},
  publisher={by R. Stepleman et al. IMACS. North-Holland}
}

@book{andrew2018reinforcement,
  author       = {Richard S. Sutton and
                  Andrew G. Barto},
  title        = {Reinforcement learning - an introduction, 2nd Edition},
  publisher    = {{MIT} Press},
  year         = {2018},
  url          = {http://www.incompleteideas.net/book/the-book-2nd.html},
  bibsource    = {dblp computer science bibliography, https://dblp.org}
}

@inproceedings{yadati2019hypergraph,
  author       = {Naganand Yadati and
                  Madhav Nimishakavi and
                  Prateek Yadav and
                  Vikram Nitin and
                  Anand Louis and
                  Partha P. Talukdar},
  editor       = {Hanna M. Wallach and
                  Hugo Larochelle and
                  Alina Beygelzimer and
                  Florence d'Alch{\'{e}}{-}Buc and
                  Emily B. Fox and
                  Roman Garnett},
  title        = {HyperGCN: {A} New Method For Training Graph Convolutional Networks
                  on Hypergraphs},
  booktitle    = {Advances in Neural Information Processing Systems 32: Annual Conference
                  on Neural Information Processing Systems 2019, NeurIPS 2019, December
                  8-14, 2019, Vancouver, BC, Canada},
  pages        = {1509--1520},
  year         = {2019},
  url          = {https://proceedings.neurips.cc/paper/2019/hash/1efa39bcaec6f3900149160693694536-Abstract.html},
  bibsource    = {dblp computer science bibliography, https://dblp.org}
}

@inproceedings{sosic2017inverse,
  author       = {Adrian Sosic and
                  Wasiur R. KhudaBukhsh and
                  Abdelhak M. Zoubir and
                  Heinz Koeppl},
  editor       = {Kate Larson and
                  Michael Winikoff and
                  Sanmay Das and
                  Edmund H. Durfee},
  title        = {Inverse Reinforcement Learning in Swarm Systems},
  booktitle    = {Proceedings of the 16th Conference on Autonomous Agents and MultiAgent
                  Systems, {AAMAS} 2017, S{\~{a}}o Paulo, Brazil, May 8-12, 2017},
  pages        = {1413--1421},
  publisher    = {{ACM}},
  year         = {2017},
  url          = {http://dl.acm.org/citation.cfm?id=3091320},
  bibsource    = {dblp computer science bibliography, https://dblp.org}
}

@article{huttenrauch2019deep,
  author       = {Maximilian H{\"{u}}ttenrauch and
                  Adrian Sosic and
                  Gerhard Neumann},
  title        = {Deep Reinforcement Learning for Swarm Systems},
  journal      = {J. Mach. Learn. Res.},
  volume       = {20},
  pages        = {54:1--54:31},
  year         = {2019},
  url          = {https://jmlr.org/papers/v20/18-476.html},
  bibsource    = {dblp computer science bibliography, https://dblp.org}
}

@inproceedings{feng2019hypergraph,
  author       = {Yifan Feng and
                  Haoxuan You and
                  Zizhao Zhang and
                  Rongrong Ji and
                  Yue Gao},
  title        = {Hypergraph Neural Networks},
  booktitle    = {The Thirty-Third {AAAI} Conference on Artificial Intelligence, {AAAI}
                  2019, The Thirty-First Innovative Applications of Artificial Intelligence
                  Conference, {IAAI} 2019, The Ninth {AAAI} Symposium on Educational
                  Advances in Artificial Intelligence, {EAAI} 2019, Honolulu, Hawaii,
                  USA, January 27 - February 1, 2019},
  pages        = {3558--3565},
  publisher    = {{AAAI} Press},
  year         = {2019},
  url          = {https://doi.org/10.1609/aaai.v33i01.33013558},
  doi          = {10.1609/AAAI.V33I01.33013558},
  bibsource    = {dblp computer science bibliography, https://dblp.org}
}

@inproceedings{kingma2014adam,
  author       = {Diederik P. Kingma and
                  Jimmy Ba},
  editor       = {Yoshua Bengio and
                  Yann LeCun},
  title        = {Adam: {A} Method for Stochastic Optimization},
  booktitle    = {3rd International Conference on Learning Representations, {ICLR} 2015,
                  San Diego, CA, USA, May 7-9, 2015, Conference Track Proceedings},
  year         = {2015},
  url          = {http://arxiv.org/abs/1412.6980},
  bibsource    = {dblp computer science bibliography, https://dblp.org}
}

@inproceedings{gasteiger2019diffusion,
  author       = {Johannes Klicpera and
                  Stefan Wei{\ss}enberger and
                  Stephan G{\"{u}}nnemann},
  editor       = {Hanna M. Wallach and
                  Hugo Larochelle and
                  Alina Beygelzimer and
                  Florence d'Alch{\'{e}}{-}Buc and
                  Emily B. Fox and
                  Roman Garnett},
  title        = {Diffusion Improves Graph Learning},
  booktitle    = {Advances in Neural Information Processing Systems 32: Annual Conference
                  on Neural Information Processing Systems 2019, NeurIPS 2019, December
                  8-14, 2019, Vancouver, BC, Canada},
  pages        = {13333--13345},
  year         = {2019},
  url          = {https://proceedings.neurips.cc/paper/2019/hash/23c894276a2c5a16470e6a31f4618d73-Abstract.html},
  bibsource    = {dblp computer science bibliography, https://dblp.org}
}

@article{brenier1991polar,
  title={Polar factorization and monotone rearrangement of vector-valued functions},
  author={Brenier, Yann},
  journal={Communications on pure and applied mathematics},
  volume={44},
  number={4},
  pages={375--417},
  year={1991},
  publisher={Wiley Online Library}
}

@article{schulman2017proximal,
  author       = {John Schulman and
                  Filip Wolski and
                  Prafulla Dhariwal and
                  Alec Radford and
                  Oleg Klimov},
  title        = {Proximal Policy Optimization Algorithms},
  journal      = {CoRR},
  volume       = {abs/1707.06347},
  year         = {2017},
  url          = {http://arxiv.org/abs/1707.06347},
  eprinttype   = {arXiv},
  eprint       = {1707.06347},
  bibsource    = {dblp computer science bibliography, https://dblp.org}
}

@book{verfurth1996review,
  author       = {R{\"{u}}diger Verf{\"{u}}hrt},
  title        = {A review of a posteriori error estimation and adaptive mesh-refinement
                  techniques},
  series       = {Advances in numerical mathematics},
  publisher    = {Wiley},
  year         = {1996},
  isbn         = {978-0-471-96795-8},
  bibsource    = {dblp computer science bibliography, https://dblp.org}
}

@inproceedings{schulman2015high,
  author       = {John Schulman and
                  Philipp Moritz and
                  Sergey Levine and
                  Michael I. Jordan and
                  Pieter Abbeel},
  editor       = {Yoshua Bengio and
                  Yann LeCun},
  title        = {High-Dimensional Continuous Control Using Generalized Advantage Estimation},
  booktitle    = {4th International Conference on Learning Representations, {ICLR} 2016,
                  San Juan, Puerto Rico, May 2-4, 2016, Conference Track Proceedings},
  year         = {2016},
  url          = {http://arxiv.org/abs/1506.02438},
  bibsource    = {dblp computer science bibliography, https://dblp.org}
}
\clearpage
\appendix

\section{Mesh Feature Definitions and Metrics}
\label{app_sec:mesh_features}

\subsection{Vertex Features}
For every vertex $z_i \in \mathcal{Z}^{(n)}$ at iteration $n$, we construct a feature vector $f_{z_i}^{(n)} \in \mathbb{R}^{d_v}$ concatenating position, physical state, and boundary constraints:
$$
    f_{z_i}^{(n)} = \left[ z_i^{(n)}, u_i^{(n)} \right]^\top.
$$
Here, $z_i^{(n)} \in \mathbb{R}^d$ denotes the coordinate vector ($d \in \{2,3\}$), which is the primary input for the Diffformer's relocation predictions. The scalar $u_i^{(n)}$ represents the nodal FEM solution value.

\subsection{Element Features}
For each element $K_j \in \mathcal{T}^{(n)}$, we assemble a feature vector $f_{K_j}^{(n)} \in \mathbb{R}^{d_e}$ that aggregates geometric quality, local solution statistics, and task-specific parameters:
$$
    f_{K_j}^{(n)} = \left[ |K_j|, \bar{u}_{K_j}, \sigma_{u,K_j}, n, \alpha_c, \mathbf{s}_{K_j}^\top, \mathbf{v}_{K_j}^\top \right]^\top.
$$
The basic components include the element measure $|K_j|$ (area in 2D) and the first two statistical moments of the solution within the element: the mean $\bar{u}_{K_j}$ and standard deviation $\sigma_{u,K_j}$, computed over the element's vertex set $V(K_j)$. We also include the current adaptation step $n$ and the cost penalty coefficient $\alpha_c$. The latter is sampled from a log-uniform distribution during training to allow the network to learn a spectrum of density preferences, enabling adaptive control over the error-cost trade-off at inference time.

\subsubsection{Element Shape Metrics ($\mathbf{s}_{K_j}$)}

\textbf{2D triangular elements.} We define $\mathbf{s}_{K_j} = [\rho_{K_j}, \sin\theta_{K_j}, \cos\theta_{K_j}]^\top$. The aspect ratio is given by $\rho_{K_j} = \ell_{\max} / \ell_{\min}$, the ratio of the longest to shortest edge lengths. The orientation is derived from the principal stretch direction of the element. Let $J_{K_j} \in \mathbb{R}^{2 \times 2}$ be the Jacobian matrix defined by the edge vectors incident to a reference vertex. We perform a Singular Value Decomposition (SVD) $J_{K_j} = U\Sigma V^\top$ and extract the angle of the principal axis as $\theta_{K_j} = \arctan2(U_{1,0}, U_{0,0})$.

\subsubsection{Solution Variability Metrics ($\mathbf{v}_{K_j}$)}
\label{app_sec:metrics}

To guide refinement in regions of high error, we compute a variability vector $\mathbf{v}_{K_j} = [\|\nabla u\|, g_{\max}, g_{\text{std}}, a_{\text{align}}]^\top$. The first component, $\|\nabla u\|$, represents the Euclidean norm of the constant solution gradient inherent to the linear finite element formulation. To capture solution fluctuations across element boundaries, we calculate the normalized solution jumps $|\Delta u_e / \ell_e|$ for every edge $e \in E(K_j)$, incorporating both their maximum, $g_{\max}$, and standard deviation, $g_{\text{std}}$, into the feature vector. Finally, to detect anisotropic features, we compute a scalar alignment metric $a_{\text{align}}$ between the solution gradient and the element's principal stretch direction $\mathbf{d}_{K_j}$ (derived from the orientation $\theta_{K_j}$ in 2D):
$$
    a_{\text{align}} = \frac{\nabla u \cdot \mathbf{d}_{K_j}}{\|\nabla u\| \|\mathbf{d}_{K_j}\|}.
$$

\section{Background on Neural Networks and Reinforcement Learning}
\label{app_sec:rl_background}
This appendix section derives the Hypergraph Neural Network architecture that provides the required permutation-invariant and dual-fidelity mesh representation for the HypeR framework. For background on reinforcement learning, swarm RL, and PPO we refer the reader to \citet{freymuth2024asmrplus}.

\subsection{Hypergraph Neural Networks}
Standard graph neural networks (GNNs) typically rely on pairwise message passing along edges, which is insufficient for capturing the volumetric constraints of finite element meshes. An element $K_j$, whether a triangle or tetrahedron, is defined by a set of vertices that collectively determine physical properties such as volume, aspect ratio, and Jacobian determinants. Decomposing these volumetric relationships into simple pairwise cliques inevitably leads to a loss of topological fidelity. To capture the full structural context required for mesh adaptation, we therefore adopt Hypergraph Neural Networks (HGNNs)~\citep{feng2019hypergraph}. By operating directly on hyperedges that connect arbitrary subsets of vertices, HGNNs naturally model the element-vertex incidence structure without artificial reduction.

\subsubsection{Hypergraph Mathematical Foundations}

We formalize the mesh as a hypergraph $\mathcal{H} = (\mathcal{V}, \mathcal{E})$, comprising $N$ vertices $\mathcal{V}$ and $M$ hyperedges (elements) $\mathcal{E}$. The connectivity is encoded by the incidence matrix $\mathbf{H} \in \{0,1\}^{N \times M}$, where $H_{ij} = 1$ if vertex $v_i$ belongs to hyperedge $e_j$. To normalize signal propagation across irregular meshes—where the number of elements meeting at a vertex (valence) varies significantly—we define the vertex degree matrix $\mathbf{D}_v$ and hyperedge degree matrix $\mathbf{D}_e$ as diagonal matrices:
$$
\mathbf{D}_v = \text{diag}\left(\sum_{j=1}^M H_{ij}\right), \quad \mathbf{D}_e = \text{diag}\left(\sum_{i=1}^N H_{ij}\right).
$$
This formulation generalizes the standard adjacency matrix, allowing us to perform spectral convolutions on sets of arbitrary cardinality.

\subsubsection{Dual Embedding Hypergraph Convolution}
\label{app_sec:hconv}

Standard HGCN architectures typically treat hyperedges as static connectors, using the incidence matrix merely to propagate information between vertices (effectively $\mathbf{H}\mathbf{H}^T$). This approach is insufficient for our multi-agent framework, where elements are active agents requiring their own latent representations to compute refinement values.

We therefore introduce a \textbf{Dual Embedding} architecture that maintains distinct, evolving latent states for both vertices, $\mathbf{X}^{(l)} \in \mathbb{R}^{N \times F}$, and hyperedges, $\mathbf{E}^{(l)} \in \mathbb{R}^{M \times F}$.  The update rule $\Phi^{(l)}: (\mathbf{X}^{(l)}, \mathbf{E}^{(l)}) \rightarrow (\mathbf{X}^{(l+1)}, \mathbf{E}^{(l+1)})$ proceeds through a bidirectional message passing cycle consisting of three distinct phases:

\paragraph{1. Vertex-to-hyperedge aggregation.}
First, elements update their state by aggregating information from their constituent vertices. This step captures the geometric configuration of the element (e.g., distortion or tangle) derived from the current vertex positions. We employ a degree-normalized convolution:
$$
\mathbf{Z}_{\mathcal{E}} = \sigma\left(\mathbf{D}_e^{-1} \mathbf{H}^T \mathbf{X}^{(l)} \mathbf{W}_{\mathcal{V} \to \mathcal{E}}^{(l)}\right),
$$
where $\mathbf{W}_{\mathcal{V} \to \mathcal{E}}^{(l)}$ is a learnable weight matrix and $\sigma$ is a non-linear activation (ReLU).

\paragraph{2. Hyperedge feature integration.}
To incorporate physics-aware data that exists strictly at the element level (such as error estimator values $\mathbf{E}_{\text{attr}}$), we fuse the aggregated geometric features $\mathbf{Z}_{\mathcal{E}}$ with the element's intrinsic attributes. This fusion is performed via concatenation and a learned projection $\mathbf{W}_{\text{pool}}$:
$$
\tilde{\mathbf{E}}^{(l+1)} = \mathbf{W}_{\text{pool}}^{(l)} \left[ \mathbf{Z}_{\mathcal{E}} \; \big\| \; \mathbf{E}_{\text{attr}}^{(l)} \right].
$$
This intermediate representation $\tilde{\mathbf{E}}^{(l+1)}$ serves as the updated state for the element agents.

\paragraph{3. Hyperedge-to-vertex aggregation.}
Finally, vertices update their representations by aggregating the enhanced features of the hyperedges they belong to. This step distributes global signals—such as high error flags or refinement needs—down to the local vertices to guide the relocation policy:
$$
\mathbf{X}^{(l+1)} = \sigma\left(\mathbf{D}_v^{-1} \mathbf{H} \tilde{\mathbf{E}}^{(l+1)} \mathbf{W}_{\mathcal{E} \to \mathcal{V}}^{(l)}\right).
$$

\subsubsection{Architectural Advantages for Mesh Adaptation}

This specific formulation offers critical advantages over standard GNNs. First, the {bidirectional information flow} ensures that vertex relocation is error-aware (receiving gradients from elements) and element refinement is geometry-aware (sensing distortion from vertices). Second, the separation of embedding spaces creates a {natural multi-agent architecture}, where the final representations $\mathbf{X}^{(L)}$ and $\mathbf{E}^{(L)}$ can be fed directly into separate actor-critic heads for vertex and element agents, respectively. Finally, the explicit degree normalization ensures {scalability}, allowing the policy to generalize zero-shot to meshes with vastly different resolutions and element counts.

\section{Hyperparameters}
\label{app_sec:hyperparameters}

\paragraph{PPO parameters.}
We employ PPO with a training schedule of 400 iterations. Each iteration collects 256 environment transitions, which are subsequently used for 5 training epochs with mini-batches of size 32. The value function loss coefficient is set to 0.5, and we apply gradient norm clipping with a maximum norm of 0.5. Both policy and value function updates use a clip range of 0.2 to ensure stable learning. 
Observations are normalized using running estimates of mean and standard deviation. We configure agent-specific discount factors: $\gamma = 1.0$ for element agents to capture long-term refinement consequences, and $\gamma = 0.1$ for vertex agents to emphasize immediate relocation benefits. Advantage estimation follows the Generalized Advantage Estimation framework \citep{schulman2015high} with $\lambda = 0.95$. Individual agent advantages are computed by subtracting agent-specific value estimates from the combined local and global returns detailed in Section 3.
These training parameters are identical to those used for ASMR++ to ensure fair comparison with the baseline method.
\paragraph{Neural network parameters.}
The shared hypergraph convolutional backbone consists of 4 message passing layers, with separate parameter sets for element and vertex agents to prevent interference between their distinct learning objectives. We employ mean aggregation as the scatter-reduce operation for all hypergraph convolutions.
The Diffformer block is configured with 4 learnable attention steps, where each step incorporates 2 diffusivity iterations (repeated applications with shared parameters) following the G-Adaptivity formulation \citep{rowbottom2025gadaptivity}. The temporal discretization step size is set to $\delta\tau = 0.1$ to ensure mesh validity guarantees.
All multi-layer perceptrons (MLPs) described in Section 3, including the policy and value heads, consist of 3 fully connected layers with $\tanh$ activation functions. We optimize all network parameters using the ADAM optimizer \citep{kingma2014adam} with a learning rate of $3.0 \times 10^{-4}$ and no learning rate scheduling.
For the ASMR++ baseline, we use the identical network architecture and hyperparameters as reported in their original work \citep{freymuth2024asmrplus}.

\paragraph{Refinement parameters.}
All adaptive mesh refinement methods require control parameters to regulate the desired refinement level of the final mesh. Both HypeR and ASMR++ employ an element penalty parameter $\alpha$ that balances the cost of adding new elements against the benefit of refinement. Following the ASMR++ methodology, both methods are trained on a range of $\alpha$ values and condition their policies on this parameter, enabling adaptive mesh resolution during inference.
For evaluation, we sample 20 penalty values log-uniformly distributed over the training ranges to produce meshes with varying element densities. The traditional heuristic methods (Oracle, ZZ1, ZZ2) refine elements based on estimated error indicators using an error threshold parameter $\theta$ that specifies which elements to refine based on the ratio between their local error and the maximum element error in the mesh. For these methods, we sample 100 values of $\theta$ uniformly distributed over their respective ranges.
Table \ref{tab:refinement_params} summarizes the parameter ranges used for all methods across the four benchmark PDEs.

\begin{table}[h]
\centering
\caption{Refinement parameter ranges for all methods across benchmark PDEs.}
\label{tab:refinement_params}
\begin{tabular}{lccccc}
\toprule
\textbf{PDE} & \textbf{HypeR} & \textbf{ASMR++} & \textbf{Oracle} & \textbf{ZZ1/ZZ2} \\
 & $\alpha$ & $\alpha$ & $\theta$ & $\theta$ \\
\midrule
Poisson & $[7.0 \times 10^{-5}, 2.0 \times 10^{-2}]$ & $[7.5 \times 10^{-3}, 2.0 \times 10^{-2}]$ & $[0, 1]$ & $[1.0 \times 10^{-4}, 1]$ \\
Stokes Flow & $[7.0 \times 10^{-5}, 2.0 \times 10^{-2}]$ & $[7.5 \times 10^{-3}, 2.0 \times 10^{-2}]$ & $[0, 1]$ & $[1.0 \times 10^{-4}, 1]$ \\
Linear Elasticity & $[7.0 \times 10^{-5}, 2.0 \times 10^{-2}]$ & $[7.5 \times 10^{-3}, 2.0 \times 10^{-2}]$ & $[0, 1]$ & $[1.0 \times 10^{-4}, 1]$ \\
Heat Diffusion & $[7.0 \times 10^{-5}, 2.0 \times 10^{-2}]$ & $[7.5 \times 10^{-3}, 2.0 \times 10^{-2}]$ & $[0, 1]$ & $[1.0 \times 10^{-4}, 1]$ \\
\bottomrule
\end{tabular}
\end{table}

These parameter ranges ensure comprehensive evaluation across the spectrum of mesh densities while maintaining fair comparison conditions between all methods.

\section{Implementation details}
\label{app_sec:implementation}
We detail the mechanisms for ensuring geometric validity during stochastic exploration and the enforcement of boundary conditions via coordinate transformations.

\subsection{Mesh Tangling Detection and Penalization}

Mesh tangling occurs when vertex displacements cause elements to invert ($\det(J) \leq 0$), making the FEM system ill-posed and preventing error evaluation. For linear simplicial elements (triangles in 2D) the Jacobian is constant per element, so validity reduces to a single check per element:
$$\det(J_{K_j^{(n)}}) = \det\begin{bmatrix}
z_{i_1}^{(n)} - z_{i_0}^{(n)} \\
z_{i_2}^{(n)} - z_{i_0}^{(n)}
\end{bmatrix} > 0.$$

\subsubsection{Tangling Penalization and Recovery}

Although the diffusion-based policy guarantees $\det(J_\Phi) > 0$ for its deterministic mean $\mu_i$, PPO samples from a stochastic Gaussian $\pi_\theta^r(\cdot|H^{(n)}) = \mathcal{N}(\mu_i, \Sigma_i)$, and large deviations during early high-entropy training can cause inversions. We handle this with a \textit{geometric rollback} and targeted penalization: only the culpable vertices—those incident to inverted elements—are penalized, ensuring precise credit assignment.

\begin{figure}[t]
\centering
\includegraphics[width=0.8\textwidth]{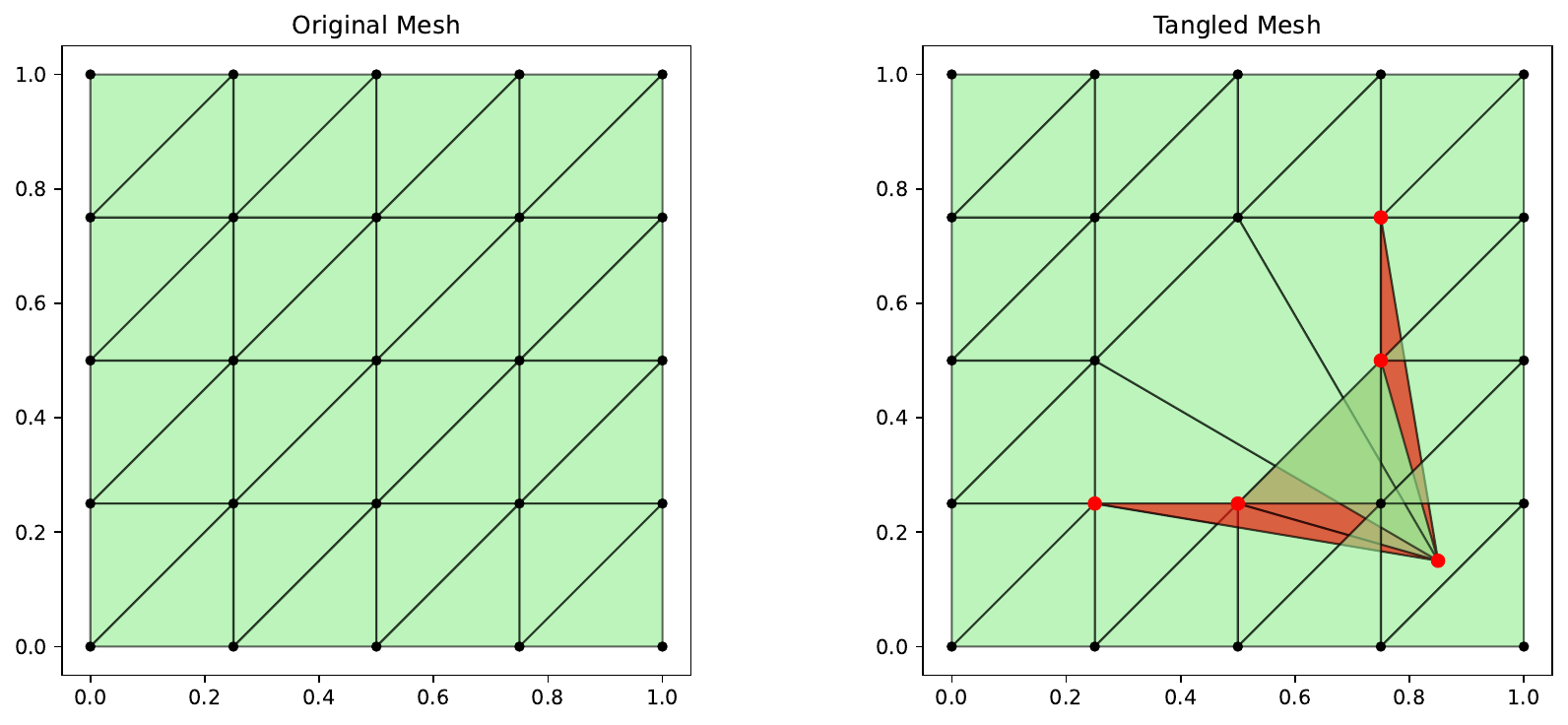}
\caption{Illustration of the local penalization scheme. A purposefully displaced central vertex (right) causes geometric inversion in the adjacent elements (red). The penalty signal $r = -1.3$ is assigned exclusively to the vertices defining these inverted elements, isolating the geometric violation to the responsible agents.}
\label{fig:mesh_tangling}
\end{figure}

Let $\mathcal{I}^{(n)} = \{K \in \mathcal{T}_{\Phi^{(n)}} : \det(J_{K}) \leq 0\}$. The penalized vertex set is:
$$
\mathcal{V}_{\text{penalty}}^{(n)} = \bigcup_{K \in \mathcal{I}^{(n)}} V(K).
$$
Vertices in $\mathcal{V}_{\text{penalty}}^{(n)}$ receive $r^{\text{loc}}_{z_i} = -\beta$ (with $\beta = 1.3$, exceeding the $[-1,1]$ reward bound), all others receive zero. The invalid step is then discarded ($Z^{(n)} \leftarrow Z^{(n-1)}$) and training continues from the previous valid geometry. This mechanism, visualized in Figure~\ref{fig:mesh_tangling}, lets agents learn the validity constraint without solver crashes or episode termination.

\subsection{Boundary Movement Constraints}
\label{app_sec:boundary_constraints}

Boundary vertices must respect the domain geometry $\Omega$. We classify them as: \emph{corner vertices} ($\mathcal{V}_{\text{corner}}$), which are fixed; and \emph{edge vertices} ($\mathcal{V}_{\text{edge}}$), which may slide along their boundary segment. Constraints are enforced structurally by masking the Diffformer's adjacency matrix so that each boundary vertex attends only to neighbors on the same geometric component $\mathcal{V}_{\partial}^{(k)}$:
$$
\tilde{A}_{ij} = \begin{cases}
A_{ij} & \text{if } i \in \mathcal{V}_{\text{interior}} \\
A_{ij} & \text{if } i, j \in \mathcal{V}_{\partial}^{(k)} \\
0 & \text{otherwise}
\end{cases}
$$
Since diffusion moves each vertex into the convex hull of its neighbors, and our domain boundaries are linear, this masking guarantees that boundary vertices remain on their assigned component—no explicit projection is needed.

\subsubsection{Boundary-Consistent Stochastic Policy}

The deterministic mean respects the mask, but PPO's stochastic exploration would violate boundary constraints if actions were sampled from an unconstrained Gaussian in $\mathbb{R}^d$. Projecting back onto the boundary distorts the PDF and invalidates the likelihood ratio. Instead, we parameterize the policy in each vertex's local DOFs space via an orthonormal basis $\mathbf{B}_i$:
\begin{itemize}
    \item Edge vertices: $\mathbf{B}_i = [\mathbf{t}_i] \in \mathbb{R}^{d \times 1}$ (unit tangent); 
    \item Corner vertices: $\mathbf{B}_i = \emptyset$.
\end{itemize}
Actions are sampled as $a_{z_i} = \mathbf{B}_i \boldsymbol{\alpha}$, where $\boldsymbol{\alpha} \sim \mathcal{N}(\mu_t, \sigma_t^2)$ for edge vertices and $\delta(\mathbf{z}_i)$ for corners. Log-probabilities for PPO are computed in the reduced space via $\boldsymbol{\alpha} = \mathbf{B}_i^T a_{z_i}$, preserving correct normalization and differentiability.

\end{document}